\documentclass[10pt,reqno]{amsart}
\usepackage{amsmath}
\usepackage{amssymb}
\usepackage{a4wide}
\usepackage{soul}

\newtheorem{thm}{Theorem}[section]
\newtheorem{cor}[thm]{Corollary}
\newtheorem{lem}[thm]{Lemma}
\newtheorem{defn}[thm]{Definition}
\newtheorem{prop}[thm]{Proposition}

\newtheorem{remark}[thm]{Remark}

\def\bea{\begin{eqnarray}}
\def\eea{\end{eqnarray}}
\def\bean{\begin{eqnarray*}}
\def\eean{\end{eqnarray*}}
\def\ds{\displaystyle}
\def\nm{\noalign{\medskip}}
\def\R{{\mathcal{R}}}
\def\Z{{\mathcal{Z}}}

\def\V{{\mathcal{V}}}
\def\K{{\mathcal{K}}}
\def\F{{\Gamma_s(\Z)}}

\def\P{{\mathcal{P}}}
\def\L{{\mathcal{L}}}

\def\S{{\mathcal{S}}}
\def\D{{\mathcal{D}}}
\newcommand{\field}[1]{\mathbb{#1}}
\newcommand{\rz}{\field{R}}

\newcommand{\nz}{\field{N}}
\def\11{{\rm 1~\hspace{-1.2ex}l} }
\def\d{{\rm{d}}}
\def\ra{{\rangle}}
\def\la{{\langle}}
\def\cqfd{{$\blacksquare$}}

\newcounter{hypo}

\newenvironment{hyp}{
 \begin{enumerate}
\setcounter{enumi}{\value{hypo}} \item}{\stepcounter{hypo} \end{enumerate}}

\makeatletter
 \@addtoreset{equation}{section}
 \makeatother
 
\title[On the classical limit of self-interacting quantum field Hamiltonians ]
{On the classical limit of self-interacting quantum field Hamiltonians  with cutoffs}

\author[Z. Ammari]{Zied Ammari}
\address{Zied Ammari,  IRMAR, (UMR CNRS 6625), Univ. Rennes I,
campus de Beaulieu, 35042 Rennes Cedex France.}
\email{zied.ammari@univ-rennes1.fr}
\author[M. Zerzeri]{Maher Zerzeri}
\address{Maher Zerzeri, LAGA, (UMR CNRS 7539), Univ. Paris 13, 99 Ave J-B. Cl\'ement, F-93430 Villetaneuse, France}
\email{zerzeri@math.univ-paris13.fr}

\keywords{Classical limit, Coherent states, QFT, Wick quantization, $P(\varphi)_2$ model.}
\subjclass[2010]{81R30, 81T10, 81Q20, 81V80}

\begin{document}

\begin{abstract}
We study, using Hepp's method, the propagation of  coherent states for a general class
of self interacting bosonic quantum field theories with spatial cutoffs.
This includes models with  non-polynomial interactions in the field variables.
We show indeed that the time evolution of coherent states, in the classical limit, is
well approximated by  time-dependent affine Bogoliubov unitary transformations.
Our analysis relies on a non-polynomial Wick quantization and a specific hypercontractive estimate.
\end{abstract}

\maketitle
\setcounter{tocdepth}{2}
\tableofcontents
\newpage

\section{Introduction}
\label{sec.intro}

In the early days of quantum mechanics Niels Bohr formulated the correspondence principle stating
that classical physics and quantum physics agree in the limit of large quantum numbers. Later
Erwin Schr\"odinger discovered  the so-called coherent states which provide a bridge between the quantum and the classical theory.
The quantum dynamics of these states are indeed closely localized around the classical trajectories although the uncertainty principle asserts that it is not
possible to find a compactly supported wave function both in the  position and the impulsion representation.
Nevertheless, coherent states are the best minimizers of the uncertainty inequality
with respect to the position and momentum observables  and hence they are the most classically localized states in the phase-space.

The physical intuition behind the coherent states and its usefulness for the classical limit was put on a firm mathematical ground by K.~Hepp in his remarkable work \cite{Hep}.
Nowadays coherent states are widely used in physics, for instance in quantum optics \cite{KS},
as well as in the mathematical literature \cite{CR}.
It is  in a certain sense an effective and yet simple tool for microlocalisation (see for instance \cite{CRR,HJ,H,W}).

It was noticed  in \cite{Hep} that the classical limit can be derived not only for
one particle Schr\"odinger dynamics but also for many-body Hamiltonians
and models of quantum field theory (see also \cite{D}). Thus, the coherent states method is also effective
for infinite dimensional phase-space analysis. However, the classical limit of quantum field theories
attracted a less attention compared  to the successful semiclassical analysis
in finite dimensions and to the fast growing subject of  mean field theory (see \cite{AB,GiVe1,RodSch} and references therein).

The purpose of the present paper is to study, through propagation of coherent states, the classical limit of self interacting Bose  field theories. We extend indeed  the result of \cite{Hep} so that
it holds true for all
coherent states, for all times and for a general class of quantum field Hamiltonians with possibly unbounded non-polynomial interactions.
We also clarify the classical field equation obtained in the limit which seems to be set inaccurately 
 in \cite{Hep}. Our results apply to the models  $(\varphi^4)_1$, $(\varphi^{2n})_1$ and more generally  $P(\varphi)_2$ boson field Hamiltonians as well as some
variant of the H{\o}egh-Krohn model (see \cite{HK1,HK2}) and some recently studied models in \cite{GP1,GP2}.
The construction of such Hamiltonians was one of the beautiful results of mathematical physics established by the late sixties (see e.g.~\cite{GJ,HS,Ro1,Se,Si}).

More precisely we show that the quantum evolution of a coherent state localized around
a point  $\varphi_0$ on the phase-space is well approximated in the classical limit
by a sequezeed  coherent states centered around  $\varphi_t$ (the classical orbit starting from
$\varphi_0$ a time $t=0$) and  deformed by a time-dependent unitary Bogoliubov  transformation.
As a consequence  the classical limit of the expectation values of
the Weyl operators on time-evolved coherent states
are the exponentials of the classical field orbit in phase space.

The classical limit can be addressed either from a dynamical point of view or a variational perspective. Here we focus on dynamical issues while  variational questions were studied in \cite{Ai3,Ar}. It is also worth mentioning that an
alternative method was developed in \cite{AmNi1}, extending Wigner (or semiclassical) measures to the infinite dimensional phase-space framework.
However, it was applied only to many-body Hamiltonians with conserved number of particles. Its
adaptation to models of quantum field theory will be considered elsewhere.

{Overview of the paper}:  In Section \ref{sec.prelim-reslt} we fix some notations
and state our main results on propagation of coherent states in the classical limit.
The proof of the main theorem (Theorem \ref{coherent}) is
presented in Section \ref{sec.coherent} where we also
establish existence of global solutions for the classical equation and study a related
time-dependent quadratic dynamic. In Section \ref{sec.wick} we introduce a specific Wick
quantization, establish an hypercontractivity type inequality and
present some models of quantum field Hamiltonians covered by
the present analysis.

\section{Preliminaries and main results}
\label{sec.prelim-reslt}

The Hamiltonians of quantum field models  can be described either
in the particle or in the wave representation. In fact, the free Bose fields Hamiltonians are simply expressed in the symmetric Fock space while the interaction is a multiplication by a measurable function
on a space $L^2(M,\mu)$ related to the representation of random Gaussian processes indexed by  real Hilbert spaces.

The general framework is as follows. Let  $\mathcal{Z}$ denote a  separable Hilbert space with a scalar product $\la \cdot,\cdot\ra$ which is anti-linear in the left
argument and with the associated norm $|z|=\sqrt{\la z,z\ra}$. We assume that $\Z$ is equipped with a
complex conjugation $\mathfrak{c}:z\mapsto \mathfrak{c}(z)$  compatible with the Hilbert structure ({\it i.e.,} $\mathfrak{c}$ is antilinear, $\mathfrak{c}\circ \mathfrak{c}(z)=z$ and
$|\mathfrak{c}(z)|=|z|$). From now on we denote
$$
\overline{z}:=\mathfrak{c}(z)\,, \quad \forall z\in\Z\,,
$$
and consider $\Z_0$ to be the real subspace of $\Z$, {\it i.e.},
\bea\label{reZ}
\Z_0:=\{z\in\Z;\ \bar z=z\}\,.
\eea
The symmetric Fock space over $\Z$ is the direct Hilbert sum
\bea\label{fock}
 \Gamma_s(\Z)=\underset{n=0}{\overset{\infty}{\oplus}}\otimes_s^n\Z \,.
\eea
A  particularly convenient dense subspace of $ \Gamma_s(\mathcal{Z})$ is the space of finite particle states given by the algebraic direct sum
\bea\label{finitepar}
\D_f=\underset{n=0}{\overset{alg}{\oplus}} \otimes_s^n\mathcal{Z}\,.
\eea
 It is well-known that $\Gamma_s(\Z)$ carries a Fock unitary representation of the Weyl
 commutation relations, namely there exists a mapping $f\mapsto W(f)$ from $\Z$ into unitary operators on
 $\Gamma_s(\Z)$ satisfying
\begin{eqnarray}
\label{eq.Weylcomm}
W(f_1) W(f_2)=e^{-\frac{i\varepsilon}{2} {\rm Im}\langle f_1, f_2\rangle} \;W(f_1+f_2), \quad \forall f_1,f_2\in \Z\,.
\end{eqnarray}
 Here $\varepsilon$ is a positive sufficiently small (semiclassical) parameter and
 ${\rm Im}\langle \cdot, \cdot\rangle$ is the imaginary part of the scalar product on $\Z$
 which is in particular a symplectic  form. The so-called Weyl operators
 $W(f)$ are given by $W(f)=e^{i\Phi_s(f)}$ for all $f\in\Z$ where
$\Phi_s(f)=\frac{1}{\sqrt{2}} (a^*(f)+a(f))$ are the Segal field operators and $a^*(\cdot),a(\cdot)$  are the $\varepsilon$-dependent creation-annihilation operators satisfying
\bean
[a(f),a^*(g)]=\varepsilon\la f, g\ra \,\11, \quad [a(f),a(g)]=0=[a^*(f),a^*(g)], \quad \forall f,g\in\Z\,.
\eean

In this framework, the coherent states are the total family of vectors in the Fock space $\Gamma_s(\Z)$ given by
\begin{eqnarray}
\label{coherent-vect}
W(-i\frac{\sqrt{2}}{\varepsilon}z)\Omega=e^{-\frac{|z|^{2}}{2\varepsilon}} \sum_{n=0}^\infty  \varepsilon^{-\frac{n}{2}}
\frac{z^{\otimes n}}{\sqrt{n!}}\,,\quad \forall z\in\Z\,,
\end{eqnarray}
where $\Omega$ is the vacuum vector (i.e.,  $\Omega=(1,0,\cdots)\in\Gamma_s(\Z)$).

The free Bose field Hamiltonian in this representation is given by the second quantized operator $\d\Gamma(A)$
defined for any self-adjoint operator $A$ on $\Z$ as
\bea
\label{dgamma}
\d\Gamma(A)_{|\otimes_s^n\Z}:=\varepsilon \sum_{i=1}^n \,\11\otimes\cdots \otimes \underbrace{A}_{i^{th} position}\otimes\cdots \otimes \11\,.
\eea
In particular the $\varepsilon$-dependent number operator is defined by
$N=\d\Gamma(\11)$.

We will sometimes use the lifting operation of an operator $A$ on $\Z$
to $\F$ given by
$\Gamma(A)_{|\otimes_s^n\Z}:=A\otimes \cdots\otimes A \,.$
For instance $\Gamma(\mathfrak{c})$ defines a conjugation on the Fock space $\Gamma_s(\Z)$.

\bigskip

It is also well-known that there exist a probability space $(M,\mathfrak{T},\mu)$ and an unitary map
$\mathcal{R}:\Gamma_s(\Z)\to L^2(M,\mu)$ such that
$\R \,\Omega=1$ and $\Z_0\ni f\mapsto \Phi(f)=\R \Phi_s(f)\R^*$
is an $\mathbb{R}$-linear mapping taking values into centered gaussian random variables on
$M$ with variance $|f|^2$ (see Theorem \ref{gaussproc}). This means that any $\Phi(f)=\R \Phi_s(f)\R^*$ is a (self-adjoint)
multiplication operator on the space $L^2(M,\mu)$ for every $f\in\Z_0$.
The mapping $\mathcal{R}$ provides an unitary equivalent Fock representation of the
Weyl commutation relations on the space $L^2(M,\mu)$ called the wave representation.
 We observe that  for any $V\in\Gamma_s(\Z)$ satisfying $\Gamma(\mathfrak{c})V=V$,
$\R(V)$ is a real-valued function belonging to $L^2(M,\mu)$. Therefore $\R(V)$ can be considered as a self-adjoint multiplication
operator on $L^2(M,\mu)$. It turns that these multiplication operators $\R(V)$ on $L^2(M,\mu)$,
when transformed back to the Fock space $\Gamma_s(\Z)$, are  (possibly non-polynomial) Wick operators
$$
F_V^{Wick}=\R^{*} (\R(V)) \R\,,
$$
with an explicit Wick symbol given by
\bea
\label{fwick}
F_V(z)= \sum_{n=0}^\infty \la\frac{(z+\bar z)^{\otimes n}}{\sqrt{n!}},V^{(n)}\ra\quad\textrm{with}\,\,
V=\underset{n=0}{\overset{\infty}{\oplus}} V^{(n)}\in\Gamma_s(\Z)\,.
\eea
Recall that $V=\underset{n=0}{\overset{\infty}{\oplus}} V^{(n)}\in\Gamma_s(\Z)$ if and only if
$\underset{n=0}{\overset{\infty}{\sum}} \Vert V^{(n)}\Vert^2_{\otimes_s^n\mathcal{Z}}<\infty$. The relation
between symbols and Wick operators is studied in details in Section \ref{sec.wick}.

We shall consider the general class of Hamiltonians given by the "sum"
\bea
\label{int-ham}
H:=\d\Gamma(A)+F_V^{Wick}\,,
\eea
where $A$ is a self-adjoint operator on $\Z$ satisfying:
\begin{hyp}\label{h1}

$\mathfrak{c}A=A\mathfrak{c}$ and $A\geq m \11$ for some  $m>0$.

\end{hyp}
The multiplication operator $\R(V)$ by a function  over $M$ on the wave representation, which
 transforms unitarily to $F_V^{Wick}$ on the Fock representation,  verifies
\begin{hyp}\label{h2}

\begin{itemize}
\item[]
$\R(V)$ is a real-valued function in $L^q(M,\mu)$ for some $q>2$ and
\item[]
$e^{-t \R(V)}\in L^1(M,\mu)$ for any  $t> 0,$
\end{itemize}

\end{hyp}
where $\R$ is the transform given by Theorem \ref{gaussproc}.
The operator $H$ depends on the parameter $\varepsilon$ and it is self-adjoint under assumptions
\ref{h1} and \ref{h2} (see Theorem \ref{segal}).

Our main result is the following theorem.
\begin{thm}
\label{coherent}
Assume \ref{h1}-\ref{h2} and that
$V\in\D(e^{\alpha \Gamma(\lambda)})$
for some $\lambda>1$ and $\alpha>0$. Let $\varphi_0\in\Z$ and $\Psi\in\D_f$ then there exists for every
 $t\in \mathbb{R}$ a  finite $\varepsilon$-independent bound $c(t,\Psi)>0$ such that the inequality
$$
\left\|e^{-i \frac{t}{\varepsilon} H} W(-i\frac{\sqrt{2}}{\varepsilon}\varphi_0)\Psi-e^{i\frac{\omega(t)}{\varepsilon}}\,
W(-i\frac{\sqrt{2}}{\varepsilon}\varphi_{t})  U_{2}(t,0)\Psi\right\|_{\F}\leq
c(t,\Psi)\;\sqrt{\varepsilon}\;,
$$
holds uniformly in  $\varepsilon>0$.
Here $\varphi_t$ is the mild  solution of the field equation
\bea
\label{fieldeq}
i\partial_t \varphi_t=A\varphi_t+\partial_{\bar z} F_V(\varphi_t)\,
\eea
with initial data $\varphi_0$, the function $\omega(t)$ is given by
$$
\omega(t)=\ds\int_0^t \sum_{k=0}^\infty \frac{(k-2)}{2} \la \frac{(\varphi_s+\bar\varphi_s)^{\otimes k}}{\sqrt{k!}}, V^{(k)}\ra\, ds\,,
$$
and $U_2(t,s)$ is the unitary propagator of a time-dependent quadratic Hamiltonian given by Corollary \ref{quad-propag}.
\end{thm}
\begin{remark}
(i) Explicitly the assumption $V\in\D(e^{\alpha \Gamma(\lambda)})$  means
 $\underset{n=0}{\overset{\infty}{\sum}} e^{2\alpha \lambda^n} ||V^{(n)}||^2 <\infty$. It is mainly due to the weak regularity properties of $U_2(t,s)$,
see Proposition \ref{prop.regU2}.\\
 (ii) Using hypercontractive estimates (see Lemma \ref{lem.hyperbound1}), the condition
 $V\in\D(e^{\alpha \Gamma(\lambda)})$ implies $\R(V)\in \underset{p\geq 2}{ \bigcap}L^p(M,\mu)$.\\
 (iii) We give an example fulfilling the  assumptions.  Let $V=\underset{n=0}{\overset{\infty}{\oplus}}
a_{2n} \varphi^{\otimes 2n}$ such that $\varphi\in\Z$, $a_m\geq 0$ and $\underset{m=0}{\overset{\infty}{\sum}}
 a_m^2 e^{2\alpha \lambda^m} |\varphi|^{2m}<\infty$ for some $\lambda>1$ and $\alpha>0$.
Then $V\in\D(e^{\alpha \Gamma(\lambda)})$ and assumption \ref{h2} is satisfied since $\R(V)\geq 0$.
 Moreover $\R(V)\notin L^\infty(M,\mu)$ if $V\neq 0$.\\
(iv) The above theorem holds with the following explicit bound for $t>0$ (with similar bound if $t<0$)
$$
c(t,\Psi)= C ||e^{\alpha\lambda^{\frac{N}{\varepsilon}}} V|| \int_0^t e^{4||\varphi_s||_\Z^2} \left[
||\sqrt{g_s(\frac{N}{\varepsilon})} \Psi||^2+ g'_s(0) \int_0^s ||V_2(r)|| dr ||\Psi||^2\right]^{\frac{1}{2}} \; ds\,,
$$
with $C>0$ depending only on $(\alpha,\lambda)$ and $V_2(r)\in\otimes_s^2\Z$ is defined by \eqref{v2}.
The functions $g_t$ and $g'_t$ are given by
 $$
  g_t(r)=\underset{k=0}{\overset{\infty}{\sum}} e^{-\alpha_0\lambda^{k}} e^{ 2\sqrt{2} \lambda_0^k \int_0^t ||V_2(s)||ds}\, (r+1)^k\quad \mbox{ and } \quad g'_t(r)=\frac{d}{dr}g_t(r)\,,
$$
for arbitrary $\lambda_0$ and $\alpha_0$ such that $1<\lambda_0<\lambda$ and $0<\alpha_0 \lambda^2<\alpha$. \\
(v) Furthermore, the quantity $V_2(r)\in\otimes_s^2\Z$ given by \eqref{v2} satisfies
$$
||V_2(r)||_{\Gamma_s(\Z)}\leq ||\left(\frac{N}{\varepsilon}\right)^4 V||_{\Gamma_s(\Z)} \, e^{4||\varphi_r||_\Z^2}\,.
$$
\end{remark}

\begin{cor}
\label{hepp}
Assume \ref{h1}-\ref{h2} and that $V\in\D(e^{\alpha \Gamma(\lambda)})$ for some $\lambda>1$ and $\alpha>0$.
We have for any $\xi\in\Z$ and $\varphi_0\in\Z$ the strong limit
$$
\ds s-\lim_{\varepsilon\to 0} W(-i\frac{\sqrt{2}}{\varepsilon}\varphi_0)^*
\, e^{i\frac{t}{\varepsilon} H} \,W(\xi)\, e^{-i\frac{t}{\varepsilon} H}\, W(-i\frac{\sqrt{2}}{\varepsilon}\varphi_0)=e^{i\sqrt{2}{\rm Re}\la\xi,\varphi_t\ra}\,\11\,,
$$
with $\varphi_t$  solving the classical field equation \eqref{fieldeq} with initial data $\varphi_0$.
\end{cor}
\noindent{\bf Proof.}
It is enough to prove the limit
\bean
\lim_{\varepsilon\to 0} \la e^{-i\frac{t}{\varepsilon} H} W(-i\frac{\sqrt{2}}{\varepsilon}\varphi_0)\,\Psi,\,W(\xi)\,
e^{-i\frac{t}{\varepsilon} H}\, W(-i\frac{\sqrt{2}}{\varepsilon}\varphi_0)\Phi\ra=e^{i\sqrt{2}{\rm Re}(\xi,\varphi_t)}\,\la \Psi,\Phi\ra\,,
\eean
for any $\Psi,\Phi\in\D_f$. Now applying Theorem \ref{coherent} for a fixed time $t$ yields
\bean
\la e^{-i\frac{t}{\varepsilon} H} W(\frac{\sqrt{2}}{i\varepsilon}\varphi_0)\Psi,W(\xi)
e^{-i\frac{t}{\varepsilon} H} W(\frac{\sqrt{2}}{i\varepsilon}\varphi_0)\Phi\ra \hspace{-.1in}&=&\hspace{-.1in}
\la  W(\frac{\sqrt{2}}{i\varepsilon}\varphi_t) U_2(t,0)\Psi,W(\xi)\, W(\frac{\sqrt{2}}{i\varepsilon}\varphi_t)U_2(t,0)\Phi\ra\\
&&+O(\sqrt{\varepsilon}).
\eean
But using the Weyl commutation relations \eqref{eq.Weylcomm} we obtain
\bean
\la  W(\frac{\sqrt{2}}{i\varepsilon}\varphi_t)\,U_2(t,0)\Psi,\,W(\xi)\, W(\frac{\sqrt{2}}{i\varepsilon}\varphi_t)\,U_2(t,0)\Phi\ra=
\la  U_2(t,0)\Psi,\,W(\xi)\, U_2(t,0)\Phi\ra e^{i\sqrt{2}{\rm Re}\la\xi,\varphi_t\ra}\,.
\eean
Thus, we obtain the claimed limit since $ s-\underset{\varepsilon\to 0}{\lim}W(\xi)=\11$ and $U_2(t,0)$ is $\varepsilon$-independent unitary operator.
\hfill \cqfd

\bigskip\noindent
{\sl Outline of the proof of Theorem \ref{coherent}}:
The proof of our main result
relies on a Taylor expansion of the Hamiltonian $H$ around the classical orbit $\varphi_t$ satisfying the field equation \eqref{fieldeq}.
Formally the Hamiltonian $H$ is a Wick quantization of the function
$$
h(z)=\la z, A z\ra+F_V(z)\,.
$$
The symbol of the translated operator $h(z+\varphi_t)^{\textrm Wick}$ of $H$ in the
 phase-space can be expanded as a sum of three terms $h(\varphi_t)$, a field operator and a time-dependent quadratic Hamiltonian, plus higher order terms on creation-annihilation operators.
 The  first and the second terms  provide an approximation for the evolution of coherent states.
More precisely to show  Theorem \ref{coherent}, we formally differentiate the quantity
$$
\mathcal{Y}(t)=e^{i\frac{t}{\varepsilon} H} e^{i\frac{\omega(t)}{\varepsilon}} W(-i\frac{\sqrt{2}}{\varepsilon}\varphi_t) U_2(t,0)\,.
$$
So, we obtain
$$
-i\varepsilon \partial_t \mathcal{Y}(t)= e^{i\frac{t}{\varepsilon} H} e^{i\frac{\omega(t)}{\varepsilon}} W(-i\frac{\sqrt{2}}{\varepsilon}\varphi_t)
\left[h(z+\varphi_t)^{Wick}-A_0(t)-\sqrt{\varepsilon} A_1(t)-\varepsilon A_2(t)\right] U_2(t,0)\,,
$$
where we have used $W(-i\frac{\sqrt{2}}{\varepsilon}\varphi_t)^* H W(-i\frac{\sqrt{2}}{\varepsilon}\varphi_t)=h(z+\varphi_t)^{Wick}$ and  $A_0(t), A_1(t), A_2(t)$ are
 $\varepsilon$-independent Wick monomials. It turns that
 $F_{R(t)}^{Wick}=h(z+\varphi_t)^{Wick}-A_0(t)-\sqrt{\varepsilon} A_1(t)-\varepsilon A_2(t)$ is a Wick operator
 of order $\varepsilon^{{\frac{3}{2}}}$. This leads to the formal estimate  for $t>0$
\bea
\label{sec2-2}
\left\|\mathcal{Y}(t)\Psi- W(-i\frac{\sqrt{2}}{\varepsilon}\varphi_0)\Psi\right\|_{\F}\leq
\varepsilon^{-1} \int_0^t \left\| F_{R(s)}^{Wick} U_2(s,0)\Psi\right\|_{\F} \,ds\,.
\eea
Hence, we get the expected estimate. However, there are several domain problems that need to be handled carefully. In particular, the regularity with respect to powers of the number operator for the propagator $U_2(t,s)$ is crucial.

\section{Wick quantization}
\label{sec.wick}
We first recall the definition of Wick monomials on the Fock space. For further information
we refer the reader to \cite{AmNi1,DeGe}. Later on, we will use the wave representation
in order to extend the Wick quantization to non-polynomial symbols.

\subsection{Polynomial Wick operators}
\bigskip

\begin{defn}
We say that a function $b:\mathcal{Z}\to \mathbb{C}$ is a continuous $(p,q)$-homogeneous polynomial in the class $\mathcal{P}_{p,q}(\Z)$ if and only if
there exists a  hermitian form $\mathfrak{Q}:\otimes^q_{\text
s}\mathcal{Z}\times\otimes^p_{\text s}\mathcal{Z} \to \mathbb C$ such that
\begin{eqnarray}
\label{herm.form1}
\exists C>0, \quad \big\vert\mathfrak{Q}(\zeta,\eta)\big\vert\leq \,C\,
\Vert \zeta\Vert_{\otimes^q_{\text
s}\mathcal{Z}} \cdot \Vert \eta\Vert_{\otimes^p_{\text
s}\mathcal{Z}},\quad \forall (\zeta,\eta)\in \otimes^q_{\text
s}\mathcal{Z}\times\otimes^p_{\text s}\mathcal{Z}\\ \label{herm.form2}
\mathfrak{Q}(\lambda\zeta,\mu\eta)=\overline{\lambda}^{q}\mu^{p}\mathfrak{Q}(\zeta,\eta),\quad
\forall (\zeta,\eta)\in \otimes^q_{\text
s}\mathcal{Z}\times\otimes^p_{\text s}\mathcal{Z}, \quad \forall \lambda,\mu\in \mathbb{C}
\end{eqnarray}
and
\begin{equation}\label{herm.form3}
b(z)=\mathfrak{Q}(z^{\otimes q},z^{\otimes p}),\quad \forall z\in\mathcal{Z}.
\end{equation}
The vector space spanned by all these polynomials will be denote by $\P$.
\end{defn}
We notice that the hermitian form $\mathfrak{Q}$ associated to $b$ in the above definition is unique by a polarization identity. Consequently,
for any $b\in\mathcal{P}_{p,q}(\Z)$ there exists a unique bounded operator $\tilde b\in\L(\otimes_s^p\Z,\otimes_s^q\Z)$ such that
\begin{equation}
\label{btilde}
b(z)=\la z^{\otimes q}, \;\tilde{b}\; z^{\otimes p}\ra\,, \quad \forall z\in\mathcal{Z}\,.
\end{equation}
Next we recall the definition of Wick quantization for symbols in $\P_{p,q}(\Z)$, $p,q\in\nz$.
The whole analysis depends on a small parameter $\varepsilon$ which we can choose sufficiently small or at least in $(0,1]$.\\
Let  $\mathcal{S}_n$ denote the orthogonal projection on the symmetric tensor product
$\otimes^n_{s}\mathcal{Z}$ given by
\begin{equation}
\mathcal{S}_n(\zeta_1\otimes\zeta_2\cdots\otimes\zeta_n)
=\frac{1}{n!}\sum_{\sigma\in \mathfrak{S}_n}\zeta_{\sigma(1)}\otimes \zeta_{\sigma(2)}\otimes
\cdots\otimes \zeta_{\sigma(n)}\,,
\end{equation}
where $\mathfrak{S}_n$ is the symmetric group of $n$ elements.
\begin{defn}
\label{wickpol}
The {\it Wick monomial} of a {\it symbol} $b\in \P_{p,q}(\Z)$ is the closure of the $\varepsilon$-dependent linear operator
 $b^{Wick}:\D_f\to\D_f\subset\Gamma_s(\Z)$ defined by
\bea
\label{sympol}
b^{Wick}_{|\otimes^n_s\Z}=1_{[p,+\infty)}(n)\frac{\sqrt{n!
(n+q-p)!}}{(n-p)!} \;\varepsilon^{\frac{p+q}{2}} \;\S_{n-p+q}\left(\tilde{b}\otimes \11^{\otimes (n-p)}\right)\,,
\eea
where $\tilde b \in\L(\otimes_s^p\Z,\otimes_s^q\Z)$ verifying \eqref{btilde}.
\end{defn}
\begin{remark}
1) For any $b\in\P_{p,q}(\Z)$ the  monomial $\bar b(z):=\overline{b(z)}$ belongs to
$\P_{q,p}(\Z)$ and the relation $\bar b^{Wick} \subset\big(b^{Wick}\big)^*$ holds.
Therefore (\ref{sympol}) defines a closable operator on $\Gamma_s(\Z)$ and in all the sequel $b^{Wick}$ denotes a closed operator.\\
2) The $\varepsilon$-dependent annihilation-creation operators can be written as
\begin{eqnarray*}
a^*(f)=\la z,f\ra^{Wick}   \hspace{.1in} \,, \hspace{.1in}   a(f)=\la f, z\ra^{Wick} \,.
\end{eqnarray*}
3) The Wick operator $\la z,z\ra^{Wick}$ is the number operator and more generally $
\d\Gamma(A)=\la z,A z\ra^{Wick}$.
\end{remark}
The composition of two Wick polynomials with symbols in $\P_{p,q}(\Z)$ is meaningful in the subspace $\D_f$.
In fact, one can show that for any $b_i\in\P$, ($i=1,2$), there exists a unique $c\in\P$ such that
\begin{eqnarray}
\label{eq.wickproduct}
&&
{b_1^{Wick} \; b_2^{Wick}}_{|\D_f}={c^{Wick}}_{|\D_f}\,.
\end{eqnarray}
The explicit formula of composition is presented in \cite[Proposition 2.7]{AmNi1}.

The Wick quantization of the real canonical variables are the so-called Segal field operators
$\Phi_{s}(f)=\sqrt{2}({\rm Re}\la f,z\ra)^{Wick}$, which are self-adjoint.
Furthermore, we observe that for any $b\in\P$, the polynomial $z\mapsto b(e^{-it A}z)$ belongs to $\P$
with the following formula holds true
$$
{e^{i \frac{t}{\varepsilon} \d\Gamma(A)} b(\cdot)^{Wick} e^{-i  \frac{t}{\varepsilon} \d\Gamma(A)}}={\left(b(e^{-it A}\cdot )\right)^{Wick}}\,.
$$
We recall the standard number estimate (see {\it e.g.}, \cite[Lemma 2.5]{AmNi1}). Uniformly  in $\varepsilon>0$, the inequality
\bea
\label{number_est}
\left|\la \Psi,b^{Wick}\Phi\ra\right|\leq ||\tilde b||_{\L(\otimes_s^p\Z,\otimes_s^q\Z)}\;||\la N\ra^{\frac{q}{2}}\Psi|| \times ||\la N\ra^{\frac{p}{2}}\Phi||\,,
\eea
holds for any $b\in\P_{p,q}(\Z)$.\\
We set
\bea
\label{dzero}
\D_c:={\rm vect}\{W(\varphi)\Omega;\,\varphi\in\Z_0\}\,.
\eea
\begin{lem}
 \label{dense}
The subspace $\D_c$ is dense in  the symmetric Fock space $\Gamma_s(\Z)$.
\end{lem}
\noindent
{\bf Proof.} Let $\Psi=\{\Psi^{(n)}\}_{n\geq 0}$ be a vector in $\Gamma_s(\Z)$ orthogonal to the set $\D_c$. In particular, we have
$ \la \Psi,W(\lambda \varphi)\Omega\ra_{\Gamma_s(\Z)}=0$ for any $\lambda\in\rz $ and $\varphi\in\Z_{0}$. An explicit computation yields
\bean
\la \Psi,W(\lambda \varphi)\Omega\ra_{\Gamma_s(\Z)}= e^{-\frac{\varepsilon}{4}\lambda^2||\varphi||^2}
\, \sum_{n=0}^\infty i^n \,\varepsilon^{\frac{n}{2}} \,\;
\la \Psi^{(n)},\varphi^{\otimes n}\ra\,\frac{\lambda^n}{\sqrt{2^n n!}}\,,
\eean
and hence the function $\lambda\mapsto \la \Psi,W(\lambda \varphi)\Omega\ra_{\Gamma_s(\Z)}$ is real-analytic. It follows that
$\la \Psi^{(n)},\varphi^{\otimes n}\ra=0$ for all $n\in\nz$ and $\varphi\in\Z_0$.
Since  the set $\{\varphi^{\otimes n}, \varphi\in\Z_0\}$ is total in $\otimes_s^n\Z$ for all $n\in\mathbb{N}$,
we conclude that $\Psi=0$.
\hfill\cqfd

\begin{lem}
\label{docore}
For any $b\in\P$ the subspace $\D_c$ is a core for $b^{Wick}$.
\end{lem}
\noindent
{\bf Proof.}
It is enough to show this property only for Wick monomials and with $\varepsilon=1$.
Recall that the subspace
$\mathcal{G}_0:={\rm Vect}\{W(f)\Psi, \Psi\in\D_f,\, f\in\Z\}$
 is a core for $b^{Wick}$ (see \cite[Proposition 2.10]{AmNi1}) and it contains $\D_c$. The Wick identity
\bean
\left[\la z+\bar z,\varphi\ra^n\right]^{Wick}=\sum_{r=0}^{[\frac{n}{2}]} (-1)^r \frac{n!}{r! (n-2r)!} \, ||\varphi||^{2r}\;
\left[\la z+\bar z,\varphi\ra^{Wick}\right]^{n-2r} \,,
\eean
holds true for any $\varphi\in\Z_{0}$. In particular, we have
\bea
\label{eq.1}
\varphi^{\otimes n}&=&\left[\la \frac{(z+\bar z)^{\otimes n}}{\sqrt{n!}},\varphi\ra\right]^{Wick}\Omega \nonumber\\
 &=& \sum_{r=0}^{[\frac{n}{2}]} (-1)^r \frac{\sqrt{n!}}{r! (n-2r)!} \, ||\varphi||^{2r}\;
\left[\la z+\bar z,\varphi\ra^{Wick}\right]^{n-2r}\Omega \nonumber\\ \nm
&=&\ds\lim_{t\to 0}\;\;\sum_{r=0}^{[\frac{n}{2}]} \, (-1)^r \frac{\sqrt{n!}}{r! (n-2r)!} \, ||\varphi||^{2r}\;
\left[\frac{W(\sqrt{2}t\varphi)-1}{i t}\right]^{n-2r}\Omega\,,
\eea
where in the last equality we have used the fact that $\Omega$ is $C^\infty$-vector for the field operator
 $\Phi_s(\varphi)$. Hence we have at hand an explicit sequence $\varphi(t)\in\D_c$, for $t\in \mathbb R\setminus\{0\}$,
 given by (\ref{eq.1}) approximating each element of the total family $\{\varphi^{\otimes n}$, $\varphi\in\Z_0, n\in\nz\}$.
   Moreover, $\ds\lim_{t\to 0}\; b^{Wick}\;\varphi(t)=  b^{Wick} \varphi^{\otimes n}$ since $\varphi(t),\varphi^{\otimes n}\in \D(b^{Wick})$ and $b^{Wick}$
 is closed. Therefore it follows that the closure of the graph of $(b^{Wick})_{|\D_c}$ contains the graph of
$b^{Wick}$. \hfill\cqfd

\subsection{Non-polynomial Wick operators}
\label{wave-rep}
We set
\bea
\label{vdez}
 \displaystyle\K:=\left\{F:\Z\to \mathbb{C};  \,\exists \,V=\underset{n=0}{\overset{\infty}{\oplus}} V^{(n)}\in\Gamma_s(\Z)\,; \, F(z)=
\sum_{n=0}^{\infty} \; \la \frac{(z+\bar z)^{\otimes n}}{\sqrt{n!}}, V^{(n)}\ra\,\right\}\,.
\eea
The mapping
\bean
\Xi:\Gamma_s(\Z)&\longrightarrow& \K\\
V=\underset{n=0}{\overset{\infty}{\oplus}} V^{(n)}&\longmapsto &F_V(z):=\sum_{n=0}^{\infty} \; \la \frac{(z+\bar z)^{\otimes n}}{\sqrt{n!}}, V^{(n)}\ra\,,
\eean
defines a Hilbert spaces isomorphism between $\Gamma_s(\Z)$ and $\K$ when the latter is endowed with the scalar product
\bean
\la F_{V_1},F_{V_2}\ra_{\K}:=\la V_1,V_2\ra_{\Gamma_s(\Z)}\,.
\eean
Moreover, we notice that $(\K,\la\cdot,\cdot\ra_\K)$ is a reproducing kernel Hilbert space with the explicit kernel  $K(z,w):=e^{\la z+\bar z,w+\bar w\ra}$ satisfying the pointwise relation
$$
\la K(w,\cdot),F_V(\cdot)\ra_{\K}=F_V(w)
$$
for all $F_V\in\K$ and $w\in\Z$.\\
Below we give the definition of Wick operators with symbols in the class $\K$.

\begin{defn}
\label{wickana}
The Wick operator with symbol $F_V$ in $\K$ is the closure of the $\varepsilon$-dependent linear operator defined by
\bea
\label{symana}
 F_V^{Wick} \left( W(\varphi)\Omega\right)=W(\varphi) \Gamma(\sqrt{\varepsilon})V,
 \quad \forall\varphi\in\Z_0.
\eea
\end{defn}
\noindent
We observe that for any collection $\varphi_i$, $i=1,\cdots,n$ of distinct elements of $\Z_0$
if $\underset{i=1}{\overset{n}{\sum}}\lambda_i W(\varphi_i)\Omega=0$  then $\lambda_i=0$ for $i=1,\cdots,n$. This implies
that  $F_V^{Wick}$ is a well-defined linear operator on $\D_c$.
\begin{remark}
1) Since $\D_c$ is dense in $\Gamma_s(\Z)$ the operator $F_V^{Wick}$, $F_V\in\K$, is  densely defined.\\
2) Notice that $(\overline{F_V})^{Wick}_{|\D_c}\subset (F_V^{Wick})^*$ then the operator given by (\ref{symana})  is closable.\\
3) The Wick quantization procedure of Definition \eqref{wickpol} and \eqref{wickana} coincide for symbols in $\P\cap\K$. \\
4) The classes $\K$ and $\P$ are different. In fact $|z|^2$ belongs to $\P$ but not to $\K$ and $F_{W(\psi)\Omega}\in\K$ for $\psi\neq 0$ and not in $\P$.
\end{remark}
The Wick quantization procedure given above have the following further property.
\begin{lem}
\label{com}
If $F_V\in\K$ then for all $\varphi\in\Z_0$
\bean
W(\varphi)\D(F_V^{Wick})=\D(F_V^{Wick}) \quad \mbox{ and } \quad W(\varphi)^*\, F_V^{Wick}\,W(\varphi)=F_V^{Wick}\,.
\eean
\end{lem}
\noindent{\bf Proof.}
By Definition \ref{wickana} and using the fact that $W(\varphi)\D_c\subset\D_c$, we verify that
\bea
\label{eq.2}
F_V^{Wick}\,W(\varphi)_{|\D_c}=W(\varphi)\,F_V^{Wick} \,_{|\D_c},\quad \mbox{ for all } \,\,\varphi\in\Z_0\,.
\eea
Since $\D_c$ is a core for $F_V^{Wick}$, we see that $W(\varphi)\D(F_V^{Wick})\subset \D(F_V^{Wick})$. Now the fact that $W(\varphi)$ is unitary with $W(\varphi)^*=W(-\varphi)$, $\varphi\in\Z_0$, yields the equality. Hence (\ref{eq.2}) extends to the domain
of $F_V^{Wick}$. \hfill\cqfd

\bigskip

The Wick quantization of symbols in $\K$ gives multiplication operators in the wave representation. Therefore it is convenient to switch to such representation when it is advantageous. For reader's convenience, we briefly recall some facts about  the wave representation (see \cite{DeGe,Si}).\\
Let $(M,\mathfrak{T},\mu)$ be a probability space. A random variable $X:M\to\rz$ with a finite variance $\sigma^2\geq 0$ is called
centered gaussian if and only if its characteristic function is
\bean
\int_{M} e^{-it X} \mu=e^{-\frac{1}{2}\sigma^2 t^2}\,, \quad t\in\rz\,.
\eean
Let $\mathfrak{H}$ be a real Hilbert space. A {\it gaussian random process} indexed by $\mathfrak{H}$ is a map
$\mathfrak{H}\ni f\mapsto \Phi(f)$ into centered gaussian random variables on $M$ with variance $|f|^2$ satisfying
\bean
 \Phi(f_1)+\Phi(f_2)=\Phi(f_1+f_2) \quad \mbox{ and } \quad\lambda \Phi(f)=\Phi(\lambda f) \quad a.e.
\eean
The process is called {\it full} if $\mathfrak{T}$ is the smallest $\sigma$-algebra such that $\Phi(f), f\in\mathfrak{H}$, are measurable.

\bigskip

Let $\mathfrak{M}$ be the abelian Von Neumann algebra generated by  the Weyl operators $W(f)$, $f\in\Z_0$.
The following theorem gives the wave representation of the canonical commutation relations (see e.g. \cite[Theorem I.1]{Si}).
\begin{thm}
 \label{gaussproc}
There exist a probability measure space $(M,\mathfrak{T},\mu)$ and a unitary map $\R:\Gamma_s(\Z)\to L^2(M,\mu)$ such that
\bean
(i)\; \R \,\Omega=1, \quad\quad (ii)\;\R\, \mathfrak{M} \,\R^*=L^\infty(M,\mu), \quad \quad (iii)\; \R\, \Gamma(\mathfrak{c}) \psi=\overline{\R\psi} \,.
\eean
Moreover, the map
\bean
\Z_0\ni f\mapsto \Phi(f)=\R \Phi_s(f)\R^*\,,
\eean
is a  gaussian full random process indexed by $\Z_0$.
\end{thm}
This theorem allows to see the Wick operators with symbols in $\K$  as  multiplication operators by unbounded measurable functions when represented in the space $L^2(M,\mu)$, see the following lemma.
\begin{lem}
For any $F_V\in\K$ there exists a measurable function $\mathcal{V}\in L^2(M,\mu)$ such that
\bean
\R  F_V^{Wick}\R^*\psi=\mathcal{V}\psi,\quad \forall \psi\in\R(\D_c)\subset L^2(M,\mu)\,,
\eean
with $\V$ acting as a multiplication operator on $L^2(M,\mu)$.
\end{lem}
Thanks to such identification we obtain the following results.
\begin{lem}
\label{self}
For any real-valued $F_V\in\K$, the corresponding Wick operator $F_V^{Wick}$ is essentially self-adjoint on $\D_c$.
\end{lem}
\noindent{\bf Proof.}
This follows from the fact that  $\R  F_V^{Wick}\R^*$  is a densely defined multiplication operator by a
$\mu$-a.e.~finite real-valued function on $L^2(M,\mu)$ (see \cite[Section VIII.3]{RS}).\hfill\cqfd

In the following lemma, we prove that the set of Wick operators with symbols in $\K\cap\P$ is dense,
with respect to the strong resolvent topology, in the set of Wick operators with $\K$ symbols.

\begin{lem}
Let $F_V$ be a real-valued function in $\K$, $V=\underset{n=0}{\overset{\infty}{\oplus}} V^{(n)}\in\Gamma_s(\Z)$.
For $\kappa$ integer we set $V_\kappa=\underset{n=0}{\overset{\kappa}{\oplus}} V^{(n)}$ and
$F_{V_\kappa}(z)=\underset{n=0}{\overset{\kappa}{\sum}}\la \frac{(z+\bar z)^{\otimes n}}{\sqrt{n!}},V^{(n)}\ra.$
Then the sequence of self-adjoint Wick polynomials $F_{V_\kappa}^{Wick}$ converges to $F_V^{Wick}$ in the strong resolvent sense.
\end{lem}
\noindent{\bf Proof.}
By the above lemma we know that $F_{V_\kappa}^{Wick}$ and $F_V^{Wick}$ are self-adjoint operators with a common core $\D_c$.  Therefore, it is enough to prove that
\bea
\label{eq.3}
\lim_{\kappa\to\infty}F_{V_\kappa}^{Wick} \,\Psi= F_V^{Wick} \,\Psi\,,
\eea
for any $\Psi\in\D_c$ in order to get the strong resolvent convergence (see  \cite[Theorem VIII.25]{RS}).
Since $F_{V_\kappa}\in\K$ we can apply Lemma \ref{com} and hence obtain
\bean
F_{V_\kappa}^{Wick} W(\varphi)\Omega=  W(\varphi) \,F_{V_\kappa}^{Wick} \Omega  = W(\varphi) \,
\sum_{n=0}^\kappa \varepsilon^{\frac{n}{2}}\,V^{(n)}\,.
\eean
Taking $\kappa\to\infty$, we get $\ds\lim_{\kappa\to\infty} F_{V_\kappa}^{Wick} W(\varphi)\Omega=W(\varphi) \Gamma(\sqrt{\varepsilon})V=F_V^{Wick} W(\varphi)\Omega$.
\hfill\cqfd

\subsection{Hypercontractive estimates}
\label{hyper-subsec}
We recall the well-known hypercontractive inequality (see \cite[Theorem I.17]{Si}).
\begin{lem}
\label{lem.hyperbound1}
Let $1<p\leq q<\infty$ and $0<\alpha\leq\sqrt{\frac{p-1}{q-1}}$. Then for any $\Psi\in \Gamma_s(\Z)$,
\bea
\label{hyperbound1}
||\mathcal{R} \Gamma(\alpha)\Psi||_{L^{q}(M,\mu)}\leq ||\mathcal{R}\Psi||_{L^p(M,\mu)} \;.
\eea
\end{lem}
The following lemma provides an information on the domain of Wick operators with symbols in $\K$.
\begin{lem}
\label{estana}
Let $V\in\Gamma_s(\Z)$ and $\lambda\geq \sqrt{3}$. Then for all $\varepsilon\in(0,\frac{1}{3}]$ and $\Psi\in\D(\Gamma(\lambda)) $:
\bean
\left\|F_V^{Wick}\,\Psi\right\|_{\F}\leq  ||V||_{\Gamma_s(\Z)} \;\;\left\|\Gamma(\lambda)\;\Psi\right\|_{\Gamma_s(\Z)} \,.
\eean
\end{lem}
\noindent{\bf Proof.}
Let $F_V\in\K$, $V\in\Gamma_s(\Z)$. Using H\"older inequality, we get for any  $\Psi\in\D(\Gamma(\lambda)) $
\bean
\left\|F_V^{Wick}\,\Psi\right\|_{\F}=\left\|\mathcal{R} F_V^{Wick}\mathcal{R}^{*} \,\mathcal{R} \Psi\right\|_{L^2(M,\mu)}
&=&\left\|(\mathcal{R}\Gamma(\sqrt{\varepsilon}) V) . \,(\mathcal{R} \Psi)\right\|_{L^2(M,\mu)}\,\\&\leq&
 \|\mathcal{R}\Gamma(\sqrt{\varepsilon})V\|_{L^4(M,\mu)}\;\|\mathcal{R} \Psi\|_{L^4(M,\mu)}\,.
\eean
The hypercontractive bound of Lemma \ref{lem.hyperbound1} with $p=2$ and $q=4$ yields
\bean
\left\|F_V^{Wick}\,\Psi\right\|_{\F}&\leq&   \|\mathcal{R}V\|_{L^2(M,\mu)}\;\|\mathcal{R} \Gamma(\lambda)\Psi\|_{L^2(M,\mu)}\,\\
&\leq&   \|V\|_{\Gamma_s(\Z)}\;\|\Gamma(\lambda) \Psi\|_{\Gamma_s(\Z)}\,.
\eean
\hfill\cqfd

\begin{remark}
(i) In the case $\varepsilon\in [\frac{1}{3},1]$, we can show the inequality
$$
\left\|F_V^{Wick}\,\Psi\right\|_{\F}\leq  ||\Gamma(\sqrt{3}) \,V||_{\F} \;\;\left\|\Gamma(\sqrt{3}) \;\Psi\right\|_{\Gamma_s(\Z)} \,.
$$
(ii) A crude inequality can be easily proved using the bound $C_n^k\leq 2^n$ and without resorting to hypercontractivity. Indeed for $\alpha>2$, we can show that
\bean
\left\| F^{Wick}_V\Psi\right\|&\leq& \frac{2}{\sqrt{1-\frac{4}{\alpha^2}}}
\Vert\Gamma(\sqrt{2})V\Vert_{\F}\, \Vert\Gamma(\alpha)\Psi\Vert_{\F}\,.
\eean
\end{remark}
\begin{prop}
 \label{prop.domfin}
 Let $V=\underset{n=0}{\overset{\infty}{\oplus}} V^{(n)}\in\Gamma_s(\Z)$  and set $V_\kappa=\underset{n=0}{\overset{\kappa}{\oplus}} V^{(n)}$.
Then for $\varepsilon\in(0,\frac{1}{3}]$:\\
(i) $\D_f$ is a core for $F_V^{Wick}$.\\
(ii) For any $\Psi\in\D_f$ the sequence $(F_{V_\kappa}^{Wick}\Psi)_{\kappa\in\mathbb{N}}$ converges to $F_{V}^{Wick}\Psi$.
\end{prop}
\noindent{\bf Proof.}
(i) Since $\D_f\subset \D(\Gamma(\lambda))$ for any $\lambda>0$ we see  that $\D_f\subset \D(F_V^{Wick})$ by Lemma \ref{estana}.
The explicit formula \eqref{coherent-vect} shows that any coherent vector $W(\varphi)\Omega$, $\varphi\in\Z_0$,  belongs to $\D(\Gamma(\lambda))$. Moreover
the sequence
$$
\Psi_\kappa=e^{-\frac{|\varphi|^2}{4}} \sum_{n=0}^\kappa \frac{i^n\varepsilon^{\frac{n}{2}}}{\sqrt{2^n n!}} \,\varphi^{\otimes^n}\;\in \D_f
$$
converges to $W(\varphi)\Omega$, when $\kappa\rightarrow\infty$, with respect to the graph norm of $\Gamma(\lambda)$. Therefore,
Lemma \ref{estana} proves that $\lim_\kappa F_V^{Wick} \Psi_\kappa=F_V^{Wick}W(\varphi)\Omega$.
So that
$$
{F_V}_{|\D_c}^{Wick}\subset \overline{{F_V}_{|\D_f}^{Wick}}  \subset F_V^{Wick}\quad \mbox{ and } \quad F_V^{Wick}=\overline{{F_V}_{|\D_c}^{Wick}} \,.
$$
(ii) The inequality in Lemma  \ref{estana} yields
\bea
\label{approfv}
\left\|(F_V^{Wick}-F_{V_\kappa}^{Wick})\,\Psi\right\|_{\F}\leq  ||V-V_\kappa||_{\Gamma_s(\Z)} \;\;\left\|\Gamma(\sqrt{3})\;\Psi\right\|_{\Gamma_s(\Z)} \,.
\eea
\hfill\cqfd

A more specific inequality is  needed.
\begin{prop}
\label{lem.estana2}
Let $S(\lambda)=\underset{k=0}{\overset{\infty }{\sum}}a_k \lambda^k$ be an entire function on $\mathbb{C}$ such that
$a_k>0$ for all $k\in\mathbb{N}$. For $\lambda_1> 8e$ there exists $C>0$  such that the inequality
\bea
\label{eq.entierbound}
\left\|F_V^{Wick}\,\Psi\right\|\leq 2 ||\Gamma(\sqrt{\varepsilon})V||\; ||\Psi||+C
\left(\sum_{n=0}^\infty \frac{(\lambda_1\varepsilon)^{n}}{a_{n+2}}||V^{(n)}||_{\otimes_s^n\Z}^2\right)^{\frac{1}{2}}
\;\;\left\|\sqrt{S(\frac{N}{\varepsilon})}\;\Psi\right\| \,
\eea
holds whenever the right hand side is finite.
\end{prop}
\noindent{\bf Proof.}
For $\Psi,V\in\D_f$ we write the decomposition
$$
F_V^{Wick}\Psi=F_{V^{(0)}}^{Wick}\Psi+F_{\underset{n\geq 1}{\oplus} V^{(n)}}^{Wick}\Psi^{(0)}+
F_{\underset{n\geq 1}{\oplus} V^{(0)}}^{Wick} \,\ds\left(\underset{n\geq 1}{\oplus}\Psi^{(n)}\right)\,.
$$
The first and second term are bounded by
 $$
 ||V^{(0)}|| \,||\Psi||+||\underset{n\geq 1}{\oplus}\varepsilon^{\frac{n}{2}}V^{(n)}|| \,||\Psi^{(0)}||\leq  2 ||\Gamma(\sqrt{\varepsilon}) V||_{\Gamma_s(\Z)} ||\Psi||_{\Gamma_s(\Z)} \,.
 $$
Now we can suppose that $V^{(0)}=0$ and $\Psi^{(0)}=0$ and write a Taylor expansion
\bean
F_V^{Wick}\Psi= \sum_{n\geq 1} \frac{\ds\varepsilon^{\frac{n}{2}}}{\sqrt{n!}} \sum_{k=0}^n C_n^k \sum_{m\geq n-k}
 \frac{\sqrt{m! (m+2k-n)!}}{(m+k-n)!} \, \S_{m+2k-n} V^{(n)}_{n-k,k} \otimes \11^{(m+k-n)} \,\Psi^{(m)}\,,
\eean
with $V^{(n)}_{n-k,k}\in\L(\otimes_s^{n-k}\Z,\otimes_s^k\Z)$.
Using the bound $\frac{\sqrt{m! (m+2k-n)!}}{(m+k-n)!}\leq \sqrt{m}^{n-k} \sqrt{m+n}^k$, we get
\bean
\left\|F_V^{Wick}\,\Psi\right\|_{\F}&\leq &\sum_{n\geq 1} \frac{\ds\varepsilon^{\frac{n}{2}}}{\sqrt{n!}} \sum_{k=0}^n C_n^k \sum_{m\geq n-k}  \sqrt{m}^{n-k} \sqrt{m+n}^k
\;||V^{(n)}||\;||\Psi^{(m)}||\\
&\leq &\sum_{n\geq 1} \frac{\ds\varepsilon^{\frac{n}{2}}}{\sqrt{n!}} \;||V^{(n)}|| \sum_{m\geq 1}  (\sqrt{m}+\sqrt{m+n})^n \;||\Psi^{(m)}||\,.
\eean
Cauchy-Schwarz inequality gives
\bean
 \sum_{m\geq 1}  (\sqrt{m}+\sqrt{m+n})^n \;||\Psi^{(m)}||\leq \left(\sum_{m\geq 1} \frac{(\sqrt{m}+\sqrt{m+n})^{2n}}{S(m)}\right)^{\frac{1}{2}}
 \;\left(\sum_{m\geq 1} S(m) ||\Psi^{(m)}||^2\right)^{\frac{1}{2}}\,.
\eean
Since $S(m)\geq a_{n+2} \,m^{n+2}$ and $m+n\leq 2 nm$ for $n,m$ positive integers, we get the estimate
$$
\frac{(\sqrt{m}+\sqrt{m+n})^{2n}}{S(m)}\leq \frac{2^{3n} n^n}{a_{n+2} \,m^2}\,.
$$
Hence by Cauchy-Schwarz
\bean
\left\|F_V^{Wick}\,\Psi\right\|_{\F}&\leq &
\frac{\pi}{\sqrt{6}} \sum_{n\geq 1} \frac{\varepsilon^{\frac{n}{2}}}{\sqrt{n!}}  \frac{\sqrt{2}^{3n} n^{\frac{n}{2}}}{\sqrt{a_{n+2}}}
\;||V^{(n)}||\;||S(\frac{N}{\varepsilon})^{\frac{1}{2}}\Psi||\\
&\leq &
\frac{\pi}{\sqrt{6}} \left(\sum_{n=1}^\infty a_{n+2}^{-1} \lambda_1^n\varepsilon^{n} ||V^{(n)}||_{\otimes_s^n\Z}^2\right)^{\frac{1}{2}}
\left( \sum_{n\geq 1}  \frac{2^{3n}n^n}{\lambda_1^{n} n!}  \right)^{\frac{1}{2}}
\;\;||S(\frac{N}{\varepsilon})^{\frac{1}{2}}\Psi||\,.
\eean
Since $\lambda_1 >8 e$  the sum   $\underset{n\geq 1 }{\sum}
\frac{2^{3n}n^n}{\lambda_1^{n}n!}$ is convergent.  By Proposition \ref{prop.domfin} the inequality extends
to any $V\in \Gamma_s(\Z)$ such that $\underset{n=0}{\overset{\infty}{\sum}} a_{n+2}^{-1} \ds(\lambda_1\varepsilon)^{n}||V^{(n)}||_{\otimes_s^n\Z}^2$ is finite and  $\Psi\in\D_f$ and then to any $\Psi\in\D(\sqrt{S(\frac{N}{\varepsilon})})$.
\hfill\cqfd

\bigskip

Later we will need to shift by translation a symbol $F_V\in\K$ ({\it i.e.}, $z\mapsto F_V(z+\varphi)$, $
\varphi\in\Z$). Therefore it would be convenient if the translated symbol still in $\K$. Below, we provide a simple sufficient condition ensuring such stability.
\begin{lem}
\label{trans-sym}
Let $F_V$ be in $\K$, $V=\underset{n=0}{\overset{\infty}{\oplus}} V^{(n)}\in\Gamma_s(\Z)$ and assume that
\bea
\label{shift}
\|\Gamma(\sqrt{2}) V\|_{\Gamma_s(\Z)}=\sqrt{\sum_{n=0}^\infty 2^{n} \,||V^{(n)}||_{\otimes_s^n\Z}^2} <\infty\,.
\eea
Then for any $\varphi\in\Z$ the function $z\mapsto F_V(z+\varphi)$ belongs to $\K$. Moreover, for $\varepsilon\in(0,\frac{1}{3}]$ the following relation holds true
\bea
\label{shiftform}
W(-i\frac{\sqrt{2}}{\varepsilon}\varphi)^{*} \; F_V(\cdot)^{Wick}\;W(-i\frac{\sqrt{2}}{
\varepsilon}\varphi)=(F_V(\cdot+\varphi))^{Wick}\,.
\eea
\end{lem}
\noindent{\bf Proof.}
Let $F$ be in $\K$, $F=\Xi(V)$ and $V=\underset{n=0}{\overset{\infty}{\oplus}} V^{(n)}\in\Gamma_s(\Z)$, such that
the sequence $(V^{(n)})_{n\geq 0}$ satisfies (\ref{shift}). For $\varphi\in\Z$, we have
\bean
F(z+\varphi)&=&\sum_{n=0}^\infty \frac{1}{\sqrt{n!}}\;
\la (z+\bar z+\varphi+\bar \varphi)^{\otimes n}, V^{(n)}\ra \\
&=& \sum_{p=0}^\infty \la \frac{(z+\bar z)^{\otimes p}}{\sqrt{p!}}, \sum_{n=p}^\infty \sqrt{\frac{n!}{p!}} \,
\frac{1}{(n-p)!} V^{(n)}_p\ra\,,
\eean
where $ V^{(n)}_p$ are the vectors in $\otimes_s^p\Z$ given by
\bean
V^{(n)}_p:=\S_{p} \left\langle(\varphi+\bar \varphi)^{\otimes n-p}\right|\otimes \11^{(p)}\; V^{(n)}\,.
\eean
In order to have $F(z+\varphi)\in\K$ it is enough to show that
$V_\varphi=\underset{p=0}{\overset{\infty}{\bigoplus}}\underset{n=p}{\overset{\infty}{\sum}} \sqrt{\frac{n!}{p!}} \,
\frac{1}{(n-p)!} V^{(n)}_p$ belongs to $\Gamma_s(\Z)$, (i.e.,
$\ds\sum_{p=0}^\infty\big\|\sum_{n=p}^\infty \sqrt{\frac{n!}{p!}} \,
\frac{1}{(n-p)!} V^{(n)}_p\big\|_{\otimes^p_s\Z}<\infty$). Indeed, we have by Cauchy-Schwarz inequality
\bean
\sum_{p=0}^\infty \left\|\sum_{n=p}^\infty \sqrt{\frac{n!}{p!}} \, \frac{1}{(n-p)!} V^{(n)}_p \right\|^2&\leq&
\sum_{p=0}^\infty  \left[\sum_{n=p}^\infty \sqrt{\frac{2^{n}}{(n-p)!}} \, (2||\varphi||)^{n-p} \, ||V^{(n)}||\right]^2 \\ \nm\ds
&\leq & \sum_{p=0}^\infty 2^p
  \left[\sum_{n=p}^\infty \frac{2^{n-p}}{(n-p)!}\, (2||\varphi||)^{2(n-p)}\right] \;\sum_{n=p}^\infty
  ||V^{(n)}||^2 \\\nm\ds
  &\leq& e^{8||\varphi||^2}\;\sum_{n=0}^\infty  ||V^{(n)}||^2 \left(\sum_{p=0}^n 2^{p}\right)\\ \nm\ds
  &\leq& 2   e^{8||\varphi||^2} \sum_{n=0}^\infty  2 ^n||V^{(n)}||^2\,.
\eean
Our next task is to show (\ref{shiftform}). Let $\varphi\in \Z$ and $V_\kappa=\underset{n=0}{\overset{\kappa}{\oplus}} V^{(n)}$ with $\kappa\in\nz$.
By \cite[Proposition 2.10]{AmNi1}, we know that for any $\Psi\in\D_c$
$$
W(-i\frac{\sqrt{2}}{\varepsilon}\varphi)^{*}\; F_{V_\kappa}^{Wick}\;W(-i\frac{\sqrt{2}}{
\varepsilon}\varphi)\Psi=F_{V_\kappa}(\cdot+\varphi)^{Wick}\,\Psi\,.
$$
Using the inequality \eqref{approfv}, we see that the above identity extends to $V$ instead of $V_\kappa$. Indeed, we can take the limit
$\kappa\to\infty$ in the left and right hand side since $W(-i\frac{\sqrt{2}}{
\varepsilon}\varphi)\Psi$ belongs to the domain of $\Gamma(\sqrt{3})$ without any assumption on $\varphi$.
\hfill\cqfd

\subsection{Models of quantum field theory}
\label{sub.selfadjt}
Our analysis is particulary motivated by two models of quantum field theory, namely the $P(\varphi)_2$ model and the H{\o}egh-Krohn model. The first is an
 example of scalar boson quantum field theory in two-dimensional space-time with a
self interaction given by an even positive polynomial on the neutral field with a spatial cutoff.
While the second model is less physical but it has some interest. In particular it provides an example of a non polynomial interaction.

Before presenting these two models we recall a cornerstone result in this subject which provides
the essential self-adjointness of the sum \eqref{int-ham} under some assumptions.
The final statement is the theorem below due to I.~Segal (see \cite{Se}).
It sums up several remarkable contributions  by E.~Nelson, A.~Jaffe, J.~Glimm,
L.~Rosen and many others (see e.g.~\cite{GJ,HS,Ro1,Se,Si}).
 It is also one of the beautiful results of mathematical physics which have had an impact on other fields (see \cite{BHL}).
\begin{thm}
\label{segal}
Let $A$  be a self-adjoint operator on $\Z$ satisfying \ref{h1} and
$F_V\in\K$ verifying the assumption \ref{h2}.
Then the operator
\bea
\label{ham}
H=\d\Gamma(A)+F_V^{Wick},
\eea
defined on $\D(\d\Gamma(A))\cap\D(F_V^{Wick})$ is essentially self-adjoint.
\end{thm}

\noindent
{\bf $\mathbf{P(\varphi)_2}$ model:}
Consider the following one variable real polynomial
\bean
P(x)=\sum_{j=0}^{2n}\alpha_j x^j, \;\;(\alpha_{2n}>0).
\eean
Let $\varphi(x)$ be the neutral scalar-field of mass $m_0> 0$, {\it i.e.}:
\bean
\label{scalarfield}
\varphi(x):=\int_{\rz}  e^{-ikx} \,[a^*(k)+a(-k)]\, \frac{dk}{\sqrt{\omega(k)}}\;,\;\;
\mbox{ where } \;\;\omega(k)=\sqrt{m_0^2+k^2}, \;\;\; m_0>0.
\eean
Let $g$  a nonnegative function in $L^1(\rz)\cap L^2(\rz)$ such that
$g(x)=g(-x)$. We define  $G$ as the following real-valued polynomial
\bea
\label{poly-inter}
G(z):= \sum_{j=0}^{2n} \alpha_j\,\int_{\rz}\; [\la z,\frac{e^{-ikx}}{\sqrt{\omega(k)}}\ra+\la \frac{e^{-ikx}}{\sqrt{
\omega(k)}}, z\ra]^j \,\,
g(x) \,dx, \ \textrm{for} \; z\in L^2(\rz,\frac{dk}{\sqrt{\omega(k)}}).
\eea
\begin{lem}
\label{poly-ext}
The polynomial $G$ given by (\ref{poly-inter}) has a continuous extension over $\Z$ belonging to the class $\K$.
\end{lem}
\noindent{\bf Proof.}
Let $\mathfrak{c}(z)=\overline{z(-k)}$ be a conjugation on $L^2(\mathbb{R})$. For $z\in L^2(\mathbb{R})$ with compact support we can write
\bean
G(z)= \sum_{j=0}^{2n} \alpha_j \la(z+\mathfrak{c}(z))^{\otimes j}
; \,\int_{\rz}\left(\frac{e^{-ikx}\chi(k)}{\sqrt{\omega(k)}}\right)^{\otimes j}g(x) \,dx\ra\,,
\eean
where $\chi$ is a smooth cutoff function verifying $\chi(k)z(k)=z(k)$.
One can prove that if $g\in L^1(\rz)\cap L^2(\rz)$ then $\displaystyle\int_{\rz}(\frac{e^{-ikx}}{\sqrt{\omega(k)}})^{\otimes j}g(x) \,dx$ is a symmetric function belonging to $L^2(\rz^j)$ (see \cite[Lemma 6.1]{DeGe}). This shows that
\bean
G(z)&=&\sum_{j=0}^{2n} \alpha_j \la(z+\mathfrak{c}(z))^{\otimes j}
; \,\chi(k)^{\otimes j}\int_{\rz}\left(\frac{e^{-ikx}}{\sqrt{\omega(k)}}\right)^{\otimes j}g(x) \,dx\ra\\
&=&\sum_{j=0}^{2n} \alpha_j \la(z+\mathfrak{c}(z))^{\otimes j}
; \,\int_{\rz}\left(\frac{e^{-ikx}}{\sqrt{\omega(k)}}\right)^{\otimes j}g(x) \,dx\ra\,.
\eean
\hfill\cqfd\\
The spatially cutoff Hamiltonian of self-interacting Bose fields in two dimensional space time
is  given by
\bean
H=\d\Gamma(\omega)+G(z)^{Wick}
\eean
on $\D(\d\Gamma(\omega))\cap \D(G(z)^{Wick})$. Therefore, applying Theorem \ref{segal} we see that $H$ is  essential self-adjointness on $\D(G(z)^{Wick})\cap\D(\d\Gamma(\omega))$. For further details on how to check the assumptions of Theorem \ref{segal} we refer the reader to \cite{DeGe,GJ,HS,Se}.
Thanks to Lemma \ref{poly-ext} the symbol $G(z)$ has the form
\bean
G(z)= \sum_{j=0}^{2n} \la \frac{(z+\overline{z})^{\otimes j}}{\sqrt{j!}}
; \,V^{(j)}\ra \quad \mbox{ with } \quad
V^{(j)}=\sqrt{j!} \alpha_j\int_{\rz}\left(\frac{e^{-ikx}}{\sqrt{\omega(k)}}\right)^{\otimes j}g(x) \,dx\in
\otimes_{s}^j L^2(\mathbb{R})
\eean
More general, we could consider instead of $G(z)$ a Wick symbol of the form
$$
F_V(z)=\sum_{n=0}^\infty
 \; \la \frac{(z+\bar z)^{\otimes n}}{\sqrt{n!}}, V^{(n)}\ra\,.
$$
with $V=\underset{n=0}{\overset{\infty}{\oplus}} V^{(n)}$ in the Fock space.

\bigskip
\noindent
{\bf H{\o}egh-Krohn model:}
This model is due to H{\o}egh-Krohn (see \cite{HK1,HK2}).
Let $\varphi(x)$ be the neutral scalar-field on $\mathbb{R}^d$  of mass $m_0> 0$, {\it i.e.},
\bean
\varphi(x):=\int_{\rz^d}  e^{-ipx} \,[a^*(p)+a(-p)]\, \frac{dp}{\sqrt{\omega(p)}}\;,\;\;
\mbox{ where } \;\;\omega(p)=\sqrt{m_0^2+p^2}, \;\;\; m_0>0.
\eean
Let $g$ be in $C_0^\infty(\mathbb{R}^d)$ such that $g\geq 0$, $g(x) = g(-x)$, $\int g(x)dx=1$
with support in the open ball of radius $1$ centered at the origin.
The cut-off field operator is given by
$$
\varphi_\kappa(x)=\int_{\rz^d} g_\kappa(x-y) \varphi(y) \,dy\,,\quad \mbox{ with }
\quad g_\kappa(x) =\kappa^{-d}g(\kappa^{-1} x)\,.
$$
For every $x\in\mathbb{R}^d$ the operator $\varphi_\kappa(x)$ is self-adjoint.
Let $V$ be a bounded continuous real function. We define the H{\o}egh-Krohn Hamiltonian as
\bea
\label{hoeghHam}
H=\d\Gamma(\omega)+\int _{|x|\leq r} V(\varphi_\kappa(x)) \, dx\,.
\eea
It is clearly a self-adjoint operator since the interaction is bounded. Instead of taking $V$ a bounded function we may
consider $V$ a real entire  function $V(\lambda)=\underset{n=0}{\overset{\infty}{\sum}} a_n \lambda^n$. This formally leads to the interaction
$$
\sum_{n=0}^\infty a_n \int_{|x|\leq r} (\varphi_\kappa(x))^n \, dx\,.
$$
In order to avoid possible infinities we replace $(\varphi_\kappa(x))^n$ by its normal ordering. This makes indeed the interaction well defined and so it is
given by
$$
\left(\sum_{n=0}^\infty \la (z+\mathfrak{c}(z))^{\otimes n}; a_n \int_{|x|\leq r} \left(e^{i px} \frac{\hat g_\kappa(p)}{\sqrt{\omega(p)}}\right)^{\otimes n} \,dx\ra\right)^{Wick}\,,
$$
whenever $\underset{n=0}{\overset{\infty }{\sum}} n! a_n^2 ||\frac{\hat g_\kappa}{\sqrt{\omega}}||_{L^2(\rz^d)}^2<\infty$. So that the modified Hamiltonian has the form
$$
H=\d\Gamma(\omega)+ F^{Wick}_{W},
$$
with
$$
W= \sum_{n=0}^\infty \sqrt{n!} a_n \int_{|x|\leq r} \left(e^{i px} \frac{\hat g_\kappa(p)}{\sqrt{\omega(p)}}\right)^{\otimes n} \,dx\in\Gamma_s(L^{2}(\rz^d)).
$$
The raison why we
 did not stick to the original model is that we are more interested in "analytic" perturbations on the field
 operators rather than bounded interactions. Moreover, the strategy will be different from the one employed
 here if the latter is considered.

\section{Propagation of coherent states}
\label{sec.coherent}

\subsection{Classical field equation}
\label{sec.clafi}
The classical limit relates models of quantum field theory to classical field equations. For instance the
$P(\varphi)_2$ dynamics, in the limit $\varepsilon\to 0$, leads to a nonlinear Klein-Gordon equation. In this subsection
we establish global existence and uniqueness of classical dynamics as primary information for the study of propagation of coherent states.
Although this relays on standard arguments we provide, for reader convenience, a short proof. \\
The  classical energy functional $h$  associated formally  to the quantum Hamiltonian $H$ defined in (\ref{ham})
is given by
\bea
\label{energy-func}
h(z):=\la z,Az\ra+F_V(z)\,,
\eea
for $z\in\D(A)$ and $F_V\in\K$. So that we have at hand the nonlinear evolution equation
\bean
i \partial_t \varphi= A \varphi+ \partial_{\bar z}F_V(\varphi)\,,
\eean
with initial data $\varphi_{|t=0}=\varphi_0 \in \D(A)$. In fact we only need to construct mild solutions for
\eqref{fieldeq}. So we rather focus on the integral equation associated to (\ref{fieldeq}), namely
\bea
\label{inteq}
\varphi_t=e^{-itA}\varphi_0-i\int_0^t e^{-i(t-s) A} \;\partial_{\bar z}F_V(\varphi_s) \;ds\,.
\eea
A fixed point argument shows the local existence of a unique continuous solution  in $C^0(\rz,\Z)$. Then a nonlinear Gronwall inequality allows to prove global existence.
We can also apply \cite[Theorem 1]{Re} or \cite[Theorem X.72]{RS} in order to show local existence.
\begin{thm}
\label{locex}
Let $A$ be a self-adjoint operator on $\Z$ and $F_V\in\K$. Then for any $\varphi_0\in\Z$ the integral equation \eqref{inteq} admits a unique
solution $\varphi_t$ in $C^0(\rz,\Z)$.  Moreover, the mapping $t\mapsto \tilde \varphi_t:=e^{it A}\varphi_t\in\Z$ is norm differentiable and satisfies
\bean
i\partial_t \tilde \varphi_t=\partial_{\bar z} F_V(\tilde\varphi_t)\,.
\eean
\end{thm}
\noindent{\bf Proof.}
The nonlinearity $\partial_{\bar z} F_V$ satisfies the explicit estimate
\bean
||\partial_{\bar z} F_V(\varphi)-\partial_{\bar z} F_V (\psi)||\leq 2 ||V||_{\Gamma_s(\Z)} \, \; g(\max(||\varphi||,||\psi||))\; ||\varphi-\psi||\,,
\eean
where $g(t)=\sqrt{1+\underset{n=2}{\overset{\infty}{\sum}} \frac{4^{n-2}n (n-1)}{(n-2)!} t^{2(n-2)}}$ is an increasing positive function. \\
For $T>0$, we consider on $C^0([0,T),\Z)$ the mapping
$$
{\mathcal T}(\varphi)(t)= e^{-itA}\varphi_0-i\int_0^t e^{-i(t-s) A} \;\partial_{\bar z}F_V(\varphi_s) \;ds\,.
$$
For any $\varphi$ and $\psi$ in  the closed ball $\mathcal{B}$ of radius $\alpha>0$ and centered at $e^{-itA}\varphi_0$, a direct computation yields
$$
\sup_{t\in[0,T)}||{\mathcal T}(\varphi)(t)-{\mathcal T}(\psi)(t)||\leq 2 ||V||_{\Gamma_s(\Z)} \,
T \; g(||\varphi_0||+\alpha) \sup_{t\in[0,T)} ||\varphi(t)-\psi(t)|| \,.
$$
Taking $\ds T< \alpha \frac{g(||\varphi_0||+\alpha)^{-\frac{1}{2}}}{2 ||V|| (||\varphi_0||+\alpha)} $
makes ${\mathcal T}$ a contraction on the closed ball $\mathcal{B}$ and hence it admits a unique fixe point.
This proves existence and uniqueness of local solutions for \eqref{inteq}.
A similar estimate yields
\bean
||\varphi_t||&\leq& ||\varphi_0||+\int_0^t ||\partial_{\bar z} F_V(\tilde\varphi_s)|| \,ds\\
&\leq& ||\varphi_0||+\int_0^t 2 ||V||_{\Gamma_s(\Z)} \,g(||\varphi_s||) \,ds\,,
\eean
So applying  a nonlinear Gronwall lemma, known as Bihari's inequality \cite{Bih}, we conclude that  $||\varphi_t||$ is bounded on any finite interval. So that $T^*=\infty$. \hfill \cqfd

\subsection{Time-dependent quadratic dynamics}
\label{sec.quadratic}
We consider in this subsection  the dynamics of time-dependent quadratic Hamiltonians.
This will be a steep towards the study of the semiclassical approximation of coherent states propagation.

\noindent
Let $Q_t$ be a real-valued time-dependent quadratic
polynomial given by
\bea
\label{abquad}
Q_t(z)=\frac{1}{\sqrt{2}} \la (e^{-it A} z+e^{it A}\bar  z)^{\otimes 2},w_t\ra\,,\quad t\in\rz\,,
\eea
such that the map $t\mapsto w_t\in\otimes_s^2\Z$ is norm continuous.
Notice that $Q_t$ is no more in $\K$ since the factor $e^{it A}$ has distorted  the symmetry of the symbol
$w_t$. However with an appropriate choice of the conjugation $\mathfrak{c_t}z:= e^{2itA}\bar z$ the symbol
 $Q_t(z)$ belongs to $\K_\mathfrak{c_t}$ with respect to $\mathfrak{c}_t$.
As a consequence we have the self-adjointness  of the operators $Q_t^{Wick}$ by Lemma \ref{self} since
$$
\Gamma(\mathfrak{c}_t) (e^{it A}\otimes e^{it A} w_t)=(e^{it A}\otimes e^{it A})w_t\,.
$$
Next we will use the Hilbert spaces,
\bean
\D_{+,k}:=\D(N^{\frac{k}{2}})\,,  \quad k\geq 1,
\eean
which are $\varepsilon$-independent vector spaces equipped with the inner product
$$
\la\Psi_1,\Psi_2\ra_{\D_{+,k}}:=\ds\sum_{n=0}^\infty (n^k+1)\;\la\Psi_1^{(n)},\Psi_2^{(n)}\ra_{\otimes_s^n\Z}\,.
$$
We  define the Hilbert space $\D_{-,k}$ as the completion of $\Gamma_s(\Z)$ with respect to the inner product
$$
\la\Psi_1,\Psi_2\ra_{\D_{-,k}}:=\ds\sum_{n=0}^\infty (n^k+1)^{-1}\;\la\Psi_1^{(n)},\Psi_2^{(n)}\ra_{\otimes_s^n\Z}.
$$
Thus, we have a Hilbert rigging
$$
\D_{+,k}\subset \F\subset \D_{-,k}\,.
$$
\begin{lem}
\label{qasum1}
Let $Q_t$ be the quadratic  polynomial given by (\ref{abquad}) such that
$t\mapsto w_t\in\otimes_s^2\Z$ is norm continuous. Then the mapping
$$
\rz\ni t\mapsto Q_t^{Wick}\in\L(\D_{+,k},\D_{-,k})
$$
is strongly continuous.
\end{lem}
\noindent{\bf Proof.}
By the number estimate (\ref{number_est}), it follows  that $Q_t^{Wick}$ is a bounded operator in
$\L(\D_{+,k},\D_{-,k})$ for each $t\in\rz$ and $k$ positive integer. More explicitly, we have
\bean
\left\|
\la \frac{(e^{-itA}z+e^{itA}\bar z)^{\otimes 2}}{\sqrt{2}}, w_t\ra^{Wick} \Psi\right\|_{\D_{-,k}}^2&\leq&
\frac{3}{2} \left[\left\|\la (e^{-itA}z)^{\otimes 2},w_t\ra^{Wick}\Psi\right\|_{\D_{-,k}}^2\right.\\
&&+\left\|\la (e^{itA}\bar z)^{\otimes 2},w_t\ra^{Wick}\Psi\right\|_{\D_{-,k}}^2
 \\
&& \left.+ 4 \left\|\la e^{-itA}z\otimes e^{itA}\bar z,w_t\ra^{Wick}\Psi\right\|_{\D_{-,k}}^2\right].
\eean
We estimate each of the three terms in the r.h.s in the same way as for the following one
\bean
||\la z^{\otimes 2},e^{itA}\otimes e^{itA}w_t\ra^{Wick}\Psi||_{\D_{-,k}}^2 &\leq&  \varepsilon^2\sum_{n=0}^\infty
\frac{(n+1) (n+2)}{((n+2)^k+1)} \, \left\|\S_{n+2} (e^{itA}\otimes e^{itA}\,w_t)\otimes \Psi^{(n)}\right\|_\F^2 \\
&\leq& \varepsilon^2\;||w_t||^2_{\otimes_s^2\Z} \,\left\|\Psi\right\|_{\D_{+,k}}^2.
\eean
Putting $\tilde w_t=e^{itA}\otimes e^{itA}w_t$, we obtain
\bean
\left\|
\la z^{\otimes 2},\tilde w_t-\tilde w_s\ra^{Wick}
\Psi\right\|_{\D_{-,k}}\leq
\varepsilon \,||w_t-w_s||_{\otimes_s^2\Z}\, ||\Psi||_{\D_{+,k}}\,.
\eean
\hfill\cqfd
\begin{lem}
\label{qasum2}
Let $Q_t$ be the  quadratic  polynomial given by (\ref{abquad}) such that
$t\mapsto w_t\in\otimes_s^2\Z$ is norm continuous. Then for any $k\geq 1$ there exist $C_k>0$ such that for any $\Psi,\Phi\in\D_f$
\bean
\left|\la (\frac{N}{\varepsilon})^k\Psi, Q_t^{Wick}\Phi\ra-\la Q_t^{Wick}\Psi,(\frac{N}{\varepsilon})^k\Phi\ra \right|\leq C_k \; \varepsilon \; ||w_t||_{\otimes_s^2\Z } \,\left\|\Psi\right\|_{\D_{+,k}}\;
\left\|\Phi\right\|_{\D_{+,k}}.
\eean
\end{lem}
\noindent{\bf Proof.}
We give the proof only for $k=1$, the case $k>1$ is similar. Actually, we will consider more carefully such an estimate in the proof of  Proposition \ref{prop.regU2}, where we need to explicit the dependence in  $k$ of the bound $C_k$. A simple computation yields
\bean
\la N\Psi, Q_t^{Wick}\Phi\ra-\la Q_t^{Wick}\Psi,N\Phi\ra \hspace{-.1in}&=&\hspace{-.1in}
\varepsilon^2\left[\sum_{n=0}^\infty \sqrt{2(n+1)(n+2)} \la \Psi^{(n+2)}, (e^{itA}\otimes e^{itA}\,w_t)\otimes \Phi^{(n)}\ra \right. \\ \nm\ds &
- &\left.\sum_{n=0}^\infty \sqrt{2 (n+1) (n+2)} \la (e^{itA}\otimes e^{itA} \,w_t)\otimes\Psi^{(n)},\Phi^{(n+2)}\ra \right].
\eean
By Cauchy-Schwarz inequality we get
$$
\left|\la N\Psi, Q_t^{Wick}\Phi\ra-\la Q_t^{Wick}\Psi,N\Phi\ra \right|\leq
2\sqrt{2}\varepsilon^2 ||w_t|| \left[\sum_{n=0}^\infty (n+1) ||\Phi^{(n)}||^2\right]^{\frac{1}{2}}
\left[\sum_{n=0}^\infty (n+1) ||\Psi^{(n)}||^2\right]^{\frac{1}{2}}.
$$
\hfill\cqfd

\bigskip
There exist several results on non-autonomous abstract
 linear Schr\"odinger equations ({see \it e.g.} \cite{Ka,Ki,Si0} and also \cite{GiVe1}). We will use a result in \cite[Corollary C.4]{AB} which is quite adapted to quadratic Hamiltonians of quantum field theory.

We say that the map $\rz\times\rz \ni (t,s)\mapsto U(t,s)$ is a unitary propagator of the non-autonomous Schr\"odinger equation
\bea
\label{abs_schrod}
\left\{
 \begin{array}[c]{l}
   i\varepsilon\partial_t u=Q_t^{Wick} u\,,\quad t\in \rz\\
   u(t=0)=u_0\in\D_{+,1}\,,
 \end{array}
\right.
\eea
if and only if\\
(a) $U(t,s)$ is unitary on $\F$,\\
(b) $U(t,t)=1$ and  $U(t,s) U(s,r)=U(t,r)$ for all $t,s,r\in \rz$,\\
(c) The map  $t\in \rz\mapsto U(t,s)$ belongs to $C^0(\rz,\L(\D_{+,1}))\cap C^{1}(\rz,\L(\D_{+,1},\D_{-,1}))$,
and satisfies
$$
i \varepsilon\partial_t U(t,s) \psi=Q_t^{Wick} U(t,s) \psi, \quad \forall \psi\in\D_{+,1}, \  \forall  t,s\in \rz.
$$
Here  $C^k(I,\mathfrak{B})$ denotes the space of $k$-continuously differentiable $\mathfrak{B}$-valued functions where
$\mathfrak{B}$ is endowed with the strong operator topology.

\begin{thm}
\label{abs_propag}
Let $Q_t$ be the quadratic  polynomial  given by (\ref{abquad})  such that:
\begin{itemize}
 \item $t\in \rz\mapsto w_t\in\otimes_s^2\Z$ is norm continuous,
 \item $\Gamma(\mathfrak{c})w_t=w_t$ for any $t\in\rz$.
\end{itemize}
Then the non-autonomous Cauchy problem (\ref{abs_schrod}) admits a unique unitary propagator $U(t,s)$.
Moreover for every $k\geq 1$ there exists $C_k>0$ such that for all  $s,t\in \rz$
\bea
\label{est-nelst}
\ds\left\|U(t,s)\right\|_{\L(\D_{+,k})}\leq \ds\;e^{\ds C_k\;\left|\int_s^t \left\|w_\tau\right\|\;\d\tau\right|}\;.
\eea
\end{thm}
\noindent{\bf Proof.} It follows by direct application of \cite[Corollary C.4]{AB} and using Lemma \ref{qasum1}-\ref{qasum2}.  \hfill\cqfd

The regularity property \eqref{est-nelst} of the propagator $U(t,s)$ contains the bound $C_k$ which we need to explicit its dependence in $k$. Actually, this can be done using  \cite[Corollary C.4]{AB}. However, we prefer to give such an inequality with a direct proof.
\begin{prop}
\label{prop.regU2}
 Assume the same hypothesis as in Theorem \ref{abs_propag}. Then for any $\lambda>1$ there exists
 $c>0$ such that for any integer $k$
$$
||(\frac{N}{\varepsilon}+1)^{\frac{k}{2}}U(t,0)\Psi||_{\F}\leq \ds
e^{\sqrt{2} k \lambda^k \left|\int_0^t \left\|w_s\right\|\;ds\right|}\,
\left[c k \left|\int_0^t \left\|w_s\right\|\;ds\right| ||\Psi||^2+||(\frac{N}{\varepsilon}+1)^{\frac{k}{2}}\Psi||_\F^2\right]^{\frac{1}{2}}.
$$
\end{prop}
\noindent{\bf Proof.}
By Theorem \ref{abs_propag} we know that $U(t,s)$ preserves the domains
$\D(N^{\frac{k}{2}})$ for any $k\geq 1$.
 Differentiating the function $u(t)=||(\frac{N}{\varepsilon}+1)^{\frac{k}{2}}U(t,0)\Psi||_{\F}^2$ for
 $\Psi\in\D_f$, we get
$$
u'(t) =\la U(t,0)\Psi, \frac{i}{\varepsilon} [Q_t^{Wick},(\frac{N}{\varepsilon}+1)^k] U(t,0)\Psi\ra\,.
$$
We decompose $Q_t^{Wick}=B_1^{Wick}+C^{Wick}+B_2^{Wick}$ with $B_1(z)=\frac{1}{\sqrt{2}}\la e^{-itA}z\otimes e^{-itA}z,w_t\ra$,
$C(z)=\sqrt{2} \la e^{-itA}z\otimes e^{itA}\bar z,w_t\ra$ and $B_2(z)=\overline{B_1(z)}$. The polynomial Wick calculus yields
$$
[Q_t^{Wick},(\frac{N}{\varepsilon}+1)^k]= \left((\frac{N}{\varepsilon}-1)^k-(\frac{N}{\varepsilon}+1)^k\right) B_1^{Wick}+B_2^{Wick} \left((\frac{N}{\varepsilon}-1)^k-(\frac{N}{\varepsilon}+1)^k\right)\,.
$$
Let $\Phi=U(t,0)\Psi$ then by explicit  computations and Cauchy-Schwarz inequality, we get
\bean
 \pm\la \Phi, \frac{i}{\varepsilon} [Q_t^{Wick},(N+1)^k] \Phi\ra &\leq&4\left|\sum_{n=2}^\infty \la\Phi^{(n)},
 [( n+1)^{k-1}+
 \cdots+( n-1)^{k-1}] \left(B_1^{Wick}\Phi\right)^{(n)}\ra\right|\\
 &\leq&4 \frac{||w_t||}{\sqrt{2}}\sum_{n=2}^\infty k(n+1)^{k}\,||\Phi^{(n)}|| \; ||\Phi^{(n-2)}||\\
 &\leq& 4k\frac{||w_t||}{\sqrt{2}}  \; \sqrt{\sum_{n=2}^\infty (n+1)^k\,||\Phi^{(n)}||^2} \;
 \sqrt{\sum_{n=0}^\infty (n+3)^k\,||\Phi^{(n-2)}||^2}\,.
 \eean
For $\lambda>0$ there exists $n_0\in\nz$ such that
$$
\sum_{n=0}^\infty (n+3)^k\,||\Phi^{(n-2)}||^2 \leq
(n_0+3) ||\Psi||^2+ \lambda^k \sum_{n=n_0+1}^\infty (n+1)^k\,||\Phi^{(n-2)}||^2\,.
$$
So that we obtain
$$
\pm u'(t) \leq 4k\lambda^k \frac{||w_t||}{\sqrt{2}}  \;u(t)+ 4k\frac{||w_t||}{\sqrt{2}} (n_0+3) ||\Psi||^2\,.
$$
The Gronwall lemma ends up the proof.
\hfill\cqfd

\bigskip

We will use the previous result with a specific choice of $w_t\in\otimes_s^2\Z$ related to the problem at hand.
Consider a symbol $F_V\in\K$  and let $\varphi_t$ be a solution of the field equation
(\ref{fieldeq}) with an initial data $\varphi_0\in \Z$. We define a real-valued  polynomial symbol $F_{V_2(t)}\in \K$ by
\bean
F_{V_2(t)}[z]&:=& \sum_{n=2}^\infty  \frac{n (n-1)}{\sqrt{n!}}\; \la (\varphi_t+\bar \varphi_t)^{\otimes (n-2)}\otimes (z+\bar z)^{\otimes 2},
V^{(n)}\ra \\ \nm\ds
&=& \,\la \frac{(z+\bar z)^{\otimes 2}}{\sqrt{2}}, V_2(t)\ra\in\P\cap\K\,.
\eean
We  check by direct computation that
\bea
\label{v2}
V_2(t)=\sqrt{2}\sum_{n=2}^\infty  \frac{n (n-1)}{\sqrt{n!}} \mathcal{S}_2\left\langle (\varphi_t+\bar \varphi_t)^{\otimes (n-2)} \right|\otimes 1^{(2)} \;
V^{(n)} \in\otimes_s^2\Z\,.
\eea
\begin{cor}
\label{quad-propag}
Let $\varphi_0\in \Z$, $F_V\in\K$ and $V_2(t)$ given by (\ref{v2}). Consider the family of polynomials
$$
F^{\mathfrak{c}_t}_{\tilde V_2(t)}(z)=
\frac{1}{\sqrt{2}} \la (e^{-it A} z+e^{it A}\bar  z)^{\otimes 2},V_2(t)\ra\,\in\K_{\mathfrak{c}_t}.
$$
Then the non-autonomous Cauchy problem
\bea
\label{tcauchy}
\left\{
 \begin{array}[c]{l}
   i\varepsilon\partial_t u=(F^{\mathfrak{c}_t}_{\tilde V_2(t)})^{Wick} u\,,\quad t\in \mathbb{R},\\
   u(t=s)=u_s\in\D_{+,1}\,,
 \end{array}
\right.
\eea
admits a unique unitary propagator $\tilde U_2(t,s)$ on $\F$. Furthermore, there exists $c>0$ depending only on $\varphi_0$ such that
\bean
||\tilde U_2(t,s)||_{\L(\D_{+,1})}\leq  \exp\big({\ds c\int^t_s dr \,||V_2(r)||_{\otimes_s^2\Z} }\big)\,.
\eean
Moreover $U_2(t,s)=e^{-i\frac{t}{\varepsilon}{\d \Gamma(A)}}\tilde U_2(t,s)e^{i\frac{s}{\varepsilon}{\d \Gamma(A)}}$ is a mild solution
of the Cauchy problem
\bea
\label{tcauchyU2}
\left\{
 \begin{array}[c]{l}
   i\varepsilon\partial_t u=({\d \Gamma(A)}+F_{V_2(t)}^{Wick}) u\,,\quad t\in \mathbb{R},\\
   u(t=s)=u_s\in\D_{+,1}\,,
 \end{array}
\right.
\eea
\end{cor}

The quantum quadratic dynamics $U_2(t,s)$ can be interpreted as a time dependent Bogoliubov transform of the Weyl commutation relations \eqref{eq.Weylcomm}.
\begin{prop}
\label{ccr}
Let $\varphi_0\in \Z$ and consider the propagator $\tilde U_2(t,0)$ given in Corollary \ref{quad-propag}. For a given  $\xi_0\in \Z$, we have
\bean
\tilde U_2(t,0) \,W(i\xi_0) \, \tilde U_2(0,t)=W(
i\beta(t,0)\xi_0)\,
\eean
where $\beta(t,0)$ is the symplectic propagator on $\Z$ solving the equation
\bea
\label{sympl-eq}
\left\{
\begin{array}{l}
i\partial_t \xi_t(x)=e^{it A} \partial_{\bar z}F_{\tilde V_2(t)}^{\mathfrak{c}_t}[\xi_t]\,,\\
\xi_{|t=0}=\xi_0\,,
\end{array}
\right.
\eea
such that $\beta(t,0)\xi_0=\xi_t$.
\end{prop}
\noindent{\bf Proof.}
The Cauchy problem (\ref{sympl-eq})  admits a unique solution  $\xi_t\in C^0(\rz,\Z)$ given by a time-ordered Dyson series since the mappings $L_t:u\mapsto e^{it A} \partial_{\bar z}F_{\tilde V_2(t)}^{\mathfrak{c}_t}[u]$ are  bounded $\rz$-linear operators on $\Z$ with the estimate
$$
||L_t(u)||\leq c ||u|| \;||V_2(t)||_{\otimes_s^2\Z}
$$
satisfied for all times. Therefore we have a well defined non-autonomous dynamical system
$\beta(t,s)$ such that $\beta(t,s)\xi_s=\xi_t$ verifying
$$
\beta(s,s)=1, \quad \beta(t,s)\beta(s,r)=\beta(t,r)\, \quad  \mbox{ for all } \quad  t,r,s\in\rz.
$$
Moreover $\beta(t,s)$ are symplectic transforms
 on $\Z$ for any $t,s\in\rz$ which can be checked by differentiating ${\rm Im}\la \beta(t,s)\xi,\beta(t,s)\eta\ra$ with respect to $t$ for $\xi,\eta\in \Z$. \\
Differentiate with respect to $t$ the quantity
\bean
\tilde U_2(0,t)\,W(-i\frac{\sqrt{2}}{\varepsilon}\xi_t) \, \tilde U_2(t,0)\,
\eean
in the sense of quadratic forms on $\D_{+,1}$, we get
\bea
\label{weyl_imp}
\begin{array}{lll}
i\varepsilon\partial_t \left[ \tilde U_2(0,t) \, W(-i\frac{\sqrt{2}}{\varepsilon} \xi_t)\,\tilde U_2(t,0)\right]
=& \tilde U_2(0,t) W(-i\frac{\sqrt{2}}{\varepsilon} \xi_t)
\left[ (F_{\tilde V_2(t)}^{\mathfrak{c}_t})^{Wick} \right. \\ \nm
&-W(-i\frac{\sqrt{2}}{\varepsilon} \xi_t)^* (F_{\tilde V_2(t)}^{\mathfrak{c}_t})^{Wick} W(-i\frac{\sqrt{2}}{\varepsilon} \xi_t) \\
&+\left.\left(
{\rm Re}\la \xi_t,i\partial_t\xi_t\ra+2
{\rm Re}\la z, i\partial_t\xi_t\ra^{Wick}\right) \right] \tilde U_2(t,s)\,.
\end{array}
\eea
Using \cite[Lemma 2.10]{AmNi1}, we see that
\bean
W(-i\frac{\sqrt{2}}{\varepsilon} \xi_t)^* (F_{\tilde V_2(t)}^{\mathfrak{c}_t})^{Wick} W(-i\frac{\sqrt{2}}{\varepsilon} \xi_t)&=&
\left(F_{\tilde V_2(t)}^{\mathfrak{c}_t} [z+\xi_t]\right)^{Wick}\,.
\eean
Hence the right hand side of (\ref{weyl_imp}) vanishes since
\bean
F_{\tilde V_2(t)}^{\mathfrak{c}_t} [z+\xi_t]-F_{\tilde V_2(t)}^{\mathfrak{c}_t}
-
{\rm Re}\la \xi_t,i\partial_t\xi_t\ra-2
{\rm Re}\la z, i\partial_t\xi_t\ra=0\,.
\eean
\hfill\cqfd

\subsection{Proof of the main result}
In this subsection we give the proof of our main result (Theorem \ref{coherent}).
Let $F_V$ be a real-valued function in $\K$ with $V\in\F$, i.e.,
$$
F_V(z)=\sum_{n=0}^\infty \la \frac{(z+\bar z)^{\otimes n}}{\sqrt{n!}}, V^{(n)}\ra\, \quad \mbox{ and }\quad
V=\underset{n=0}{\overset{\infty}{\oplus}} V^{(n)} \quad \mbox{ with } \quad V^{(n)}\in\otimes_s^n\Z\,.
$$
We consider the symbol $F^{\mathfrak{c}_t}_{V(t)}\in\K_\mathfrak{c_t}$, with respect to the conjugation $\mathfrak{c_t}z:= e^{2itA}\bar z$, obtained  from $V\in\F$ as follows
\bea
\label{vt}
F_{V(t)}^{\mathfrak{c}_t}(z)=\sum_{n=0}^\infty \la \frac{(e^{-itA}z+e^{itA}\bar z)^{\otimes n}}{\sqrt{n!}}, V^{(n)}\ra\,.
\eea
We first prove some  preliminary lemmas.
\begin{lem}
\label{lem1coh}
The map $\rz\ni t\mapsto e^{i\frac{t}{\varepsilon} H} e^{-i\frac{t}{\varepsilon} {\d \Gamma(A)}} \Psi$ is norm differentiable in $\F$ for any $\Psi\in\D(\Gamma(\lambda))$,
with $\lambda\geq\sqrt{3}$ and $\varepsilon\in(0,\frac{1}{3}]$. Moreover, the following identity holds
\bean
i\varepsilon \partial_t\;e^{i\frac{t}{\varepsilon} H} e^{-i\frac{t}{\varepsilon} {\d \Gamma(A)}} \Psi=-e^{i\frac{t}{\varepsilon} H} e^{-i\frac{t}{\varepsilon} {\d \Gamma(A)}} (F^{\mathfrak{c}_t}_{V(t)})^{Wick}\Psi\,.
\eean
\end{lem}
\noindent{\bf Proof.}
By Lemma \ref{estana}, we know that if $\lambda\geq \sqrt{3}$ then
$\D(\Gamma(\lambda))\subset \D(F_V^{Wick})$. Hence for $\Psi\in\D({\d \Gamma(A)})\cap\D(\Gamma(\lambda))$ with $\lambda\geq \sqrt{3}$, we have
\bean
-i\varepsilon \;e^{i\frac{t}{\varepsilon} H} e^{-i\frac{t}{\varepsilon} {\d \Gamma(A)}} \Psi &=&e^{i\frac{t}{\varepsilon} H}\; (H-{\d \Gamma(A)})\;
e^{-i\frac{t}{\varepsilon} {\d \Gamma(A)}}\Psi\\
&=& e^{i\frac{t}{\varepsilon} H}\;F_V^{Wick} \;e^{-i\frac{t}{\varepsilon} {\d \Gamma(A)}}\Psi\\
&=& e^{i\frac{t}{\varepsilon} H} \;e^{-i\frac{t}{\varepsilon} {\d \Gamma(A)}}
(F_{V(t)}^{\mathfrak{c}_t})^{Wick}\Psi.
\eean
The two last equalities hold using the fact that $H={\d \Gamma(A)}+F_V^{Wick}$ when restricted  to $\D({\d \Gamma(A)})\cap\D(\lambda^N)$.
Now, for $\Psi\in\D(\Gamma(\lambda))$ we take a sequence $\Psi_\kappa\in\D({\d \Gamma(A)})\cap\D(\Gamma(\lambda))$ such that $
\Psi_\kappa\to\Psi$ when $\kappa\to\infty$. Therefore, we can write
\bea
\label{eq.4}
e^{i\frac{t}{\varepsilon} H} e^{-i\frac{t}{\varepsilon} {\d \Gamma(A)}} \Psi_\kappa=\Psi_\kappa+\frac{i}{\varepsilon} \int_0^t
e^{i\frac{s}{\varepsilon} H} e^{-i\frac{s}{\varepsilon} {\d \Gamma(A)}} (F_{V(s)}^{\mathfrak{c}_t})^{Wick} \Psi_\kappa \;ds\,.
\eea
Letting $\kappa\to\infty$ that the same identity holds for $\Psi$ instead of $\Psi_\kappa$.
This yields that
$$
t\mapsto e^{i\frac{t}{\varepsilon} H} e^{-i\frac{t}{\varepsilon} {\d \Gamma(A)}} \Psi
$$
is norm differentiable in $\F$ for any $\Psi\in\D(\Gamma(\lambda))$, $\lambda\geq \sqrt{3}$.
\hfill\cqfd

\begin{lem}
\label{lem.domcomp}
For any $V\in\D(e^{\alpha\Gamma(\lambda)})$ with $\lambda>1$, $\alpha>0$ and any $\varphi\in\Z$ it holds for 
$\varepsilon\in(0,\frac{1}{3}]$:\\
(i) $
W(-i\frac{\sqrt{2}}{\varepsilon}\varphi) U_2(t,0) \D_f\subset \D(F_V^{Wick})$.\\
(ii) There exist a $(\varepsilon,V,t)$-independent constant $C>0$ such that for any
$\Psi\in\D_f$
\bean
||F_V^{Wick} U_2(t,0)\Psi||&\leq & C  ||e^{\alpha\lambda^{\frac{N}{\varepsilon}}} \Gamma(\sqrt{\varepsilon})V||_{\Gamma_s(\Z)} \left[
||\sqrt{g_t(\frac{N}{\varepsilon})} \Psi||^2+ g'_t(0) \left|\int_0^t ||V_2(s)|| ds\right| ||\Psi||^2\right]^{\frac{1}{2}}\,,
\eean
where $g_t(r)=\underset{k=0}{\overset{\infty}{\sum}} e^{-\alpha_0\lambda^{k}} e^{ 2\sqrt{2} \lambda_0^k \int_0^t ||V_2(s)||ds}\, (r+1)^k$ for $1<\lambda_0<\lambda$, $0<\alpha_0 \lambda^2<\alpha$ and $g'_t(r)=\frac{d}{dr}g_t(r)$.
\end{lem}
\noindent{\bf Proof.}
We observe that Lemma \ref{trans-sym} gives
$$
W(-i\frac{\sqrt{2}}{\varepsilon}\varphi)^*F_V^{Wick}W(-i\frac{\sqrt{2}}{\varepsilon}\varphi)
=F_{V_\varphi}^{Wick}\,,
$$
with $V_\varphi$ satisfying the inequality $||V_\varphi||\leq c ||\Gamma(\sqrt{2}) V||$. Hence $V_\varphi\in \D(e^{\beta\Gamma(\lambda)})$ with $0<\beta<\alpha$ and therefore it is enough to show
that  $U_2(t,0) \D_f\subset \D(F_V^{Wick})$ which follows by (ii).\\
Let $V,\Psi\in\D_f$ and $S(x)=\underset{k=0}{\overset{\infty}{\sum}} a_k x^k$ be the entire series with
$a_k=e^{-\alpha_0\lambda^{k}}$ such  that $\alpha_0>0$ and $\alpha_0 \lambda^2<\alpha$. Writing
\bean
||F_V^{Wick} U_2(t,0)\Psi||&\leq &||F_V^{Wick} S(\frac{N}{\varepsilon})^{-\frac{1}{2}}|| \; || S(\frac{N}{\varepsilon})^{\frac{1}{2}} U_2(t,0) \Psi||\,,
\eean
then applying Proposition \ref{prop.regU2} with $1<\lambda_0<\lambda$, we get
\bean
 || S(\frac{N}{\varepsilon})^{\frac{1}{2}} U_2(t,0) \Psi||&\leq & c\;
 \left[|| \sqrt{\sum_{k=0}^\infty a_k(t) (\frac{N}{\varepsilon}+1)^k} \Psi||^2+
 \sum_{k=0}^\infty k a_k(t) \left|\int_0^t \left\|V_2(s)\right\|ds\right| ||\Psi||^2\right]^{\frac{1}{2}}\,,
\eean
where $a_k(t)=a_k e^{\sqrt{2} k \lambda_0^k \left|\int_0^t \left\|V_2(s)\right\|\;ds\right|}$.
Since $\Psi\in\D_f$ there exists $m\in\mathbb{N}$ such that
$\Psi\in \underset{n=0}{\overset{m}{\oplus}}\otimes_s^n\Z$ and therefore the inequality
$$
|| \sqrt{\sum_{k=0}^\infty a_k(t) (\frac{N}{\varepsilon}+1)^k} \Psi||\leq \sqrt{\sum_{k=0}^\infty a_k(t) (m+1)^k} \;||\Psi||
$$
holds with a finite right hand side.  Using Proposition \ref{lem.estana2} we see that for $\lambda_1>8e$
\bean
||F_V^{Wick} S(\frac{N}{\varepsilon})^{-\frac{1}{2}}||
&\leq& C  \;\sqrt{\sum_{k\geq 0} a_{k+2}^{-1} (\lambda_1 \varepsilon)^n ||V^{(k)}||^2}
\leq C' \; ||e^{\alpha\lambda^{\frac{N}{\varepsilon}}}
\Gamma(\sqrt{\varepsilon}) V||\,.
\eean
\hfill\cqfd

 \bigskip

\noindent
{\bf Proof of Theorem \ref{coherent}:}\\
Since the main quantity to be estimated in Theorem \ref{coherent} is bounded by $2$, we can assume without restriction that  $\varepsilon$ is sufficiently  small.
Let $F^{\mathfrak{c}_t}_{\tilde V_2(t)}$ be a symbol in $\K_\mathfrak{c_t}$, with respect to the conjugation $\mathfrak{c_t}z:=e^{2itA}\bar z$, given by
\bea
\label{v2t}
F^{\mathfrak{c}_t}_{\tilde V_2(t)}&:=&
\la \frac{(e^{-itA}z+e^{itA}\bar z)^{\otimes 2}}{\sqrt{2}}, V_2(t)\ra\in\P\cap\K\,,
\eea
where  $V_2(t)$ is defined by \eqref{v2}.
For $\Psi\in\D_f$, we write
$$
\Theta(t):=e^{i\frac{t}{\varepsilon}
H} e^{i\frac{\omega(t)}{\varepsilon}}
W(-i\frac{\sqrt{2}}{\varepsilon}\varphi_{t}) U_{2}(t,0)\Psi =
e^{i\frac{t}{\varepsilon}
H} e^{-i\frac{t}{\varepsilon}
\d\Gamma(A)}e^{i\frac{\omega(t)}{\varepsilon}}
W(-i\frac{\sqrt{2}}{\varepsilon}\tilde\varphi_{t}) e^{i\frac{t}{\varepsilon}
\d\Gamma(A)} U_{2}(t,0)\Psi \,.
$$
We differentiate the above quantity with respect to time $t$. We recall the formula
\bea
\label{derweyl}
i \varepsilon \partial_t W(-i\frac{\sqrt{2}}{\varepsilon} \tilde\varphi_t)\Psi&=& W(-i\frac{\sqrt{2}}{\varepsilon} \tilde\varphi_t) \left[
{\rm Re}\la \tilde\varphi_t, i\partial_t \tilde\varphi_t\ra+2{\rm Re}\la z,i\partial_t\tilde\varphi_t\ra^{Wick}\right]\Psi\,,
\eea
where $\Psi\in\D_f$ (see \cite{AB,GiVe1}).
Since $e^{i\frac{t}{\varepsilon}{d\Gamma(A)}}$
commutes with the number operator $N$ we know, by Lemma \ref{lem.domcomp},  that the vector
$W(-i\frac{\sqrt{2}}{\varepsilon}\tilde\varphi_{t}) e^{i\frac{t}{\varepsilon}{d\Gamma(A)}} U_{2}(t,0)\Psi$ belongs to $\D(e^{\alpha\lambda^{\frac{N}{\varepsilon}}})$. Therefore, using Lemma \ref{lem1coh},
 we can differentiate $e^{i\frac{t}{\varepsilon}H} e^{-i\frac{t}{\varepsilon}{d\Gamma(A)}}$ then differentiate $e^{i\frac{\omega(t)}{\varepsilon}}W(-i\frac{\sqrt{2}}{\varepsilon}\tilde\varphi_{t})$ using \eqref{derweyl} and finally  differentiate  $e^{i\frac{t}{\varepsilon}\d\Gamma(A)} U_{2}(t,0)\Psi$ using Corollary \ref{quad-propag}. So that, we get
\bean
\Theta'(t)&=&\frac{i}{\varepsilon} \,e^{i\frac{t}{\varepsilon}
H} e^{-i\frac{t}{\varepsilon}
\d\Gamma(A)}e^{i\frac{\omega(t)}{\varepsilon}}
W(-i\frac{\sqrt{2}}{\varepsilon}\tilde\varphi_{t}) \left[W(-i\frac{\sqrt{2}}{\varepsilon}\tilde\varphi_{t})^*
(F^{\mathfrak{c}_t}_{V(t)})^{Wick}W(-i\frac{\sqrt{2}}{\varepsilon}\tilde\varphi_{t}) \right.\\
&& \left.+\partial_t\omega(t)-
{\rm Re}\la \tilde\varphi_t, i\partial_t \tilde\varphi_t\ra-2{\rm Re}\la z,i\partial_t\tilde\varphi_t\ra^{Wick}-(F^{\mathfrak{c}_t}_{\tilde V_2(t)})^{Wick}\right]
 e^{i\frac{t}{\varepsilon}
\d\Gamma(A)} U_{2}(t,0)\Psi\\
&=&\frac{i}{\varepsilon} \,e^{i\frac{t}{\varepsilon}
H} e^{i\frac{\omega(t)}{\varepsilon}}
W(-i\frac{\sqrt{2}}{\varepsilon}\varphi_{t}) \left[W(-i\frac{\sqrt{2}}{\varepsilon}\varphi_{t})^*
F_{V}^{Wick}W(-i\frac{\sqrt{2}}{\varepsilon}\varphi_{t})+\partial_t\omega(t) \right.\\
&& \left.-
{\rm Re}\la\varphi_t, \partial_{\bar z}F_V(\varphi_t)\ra-2{\rm Re}\la z,\partial_{\bar z}F_V(\varphi_t)\ra^{Wick}-F_{ V_2(t)}^{Wick}\right]
 U_{2}(t,0)\Psi
\eean
where $F^{\mathfrak{c}_t}_{V(t)}$ and $F^{\mathfrak{c}_t}_{\tilde V_2(t)}$ are given respectively
 by \eqref{vt}-\eqref{v2t}.
By Lemma \ref{trans-sym} we know that
$$
W(-i\frac{\sqrt{2}}{\varepsilon}\varphi_{t})^*F_{V}^{Wick} W(-i\frac{\sqrt{2}}{\varepsilon}\varphi_{t})=F_{V}(.+\varphi_t)^{Wick}.
$$
We define for  any $n,k\in\nz$, $ k\geq n$
\bean
V^{(n)}_k:=\S_n \left\langle (\varphi_t+\overline{\varphi}_t)^{\otimes (k-n)}\right|\otimes \11^{(n)} \;V^{(k)} \in\otimes_s^n\Z\,.
\eean
One can check by direct computation that $\underset{n=0}{\overset{\infty}{\sum}}\;\underset{k=n}{\overset{\infty}{\sum}}
\sqrt{\frac{k!}{n!}}\frac{1}{(k-n)!} V^{(n)}_k \in\F$.
Hence expanding
$F_{V}(.+\varphi_t)$ around $\varphi_t$ we obtain
\bean
F_{V}(z+\varphi_t)&=&\sum_{n=0}^\infty \la \frac{(z+\bar z)^{\otimes n}}{\sqrt{n!}}, \sum_{k=n}^{\infty}
\sqrt{\frac{n!}{k!}} C_k^n V^{(n)}_k\ra\,.
\eean
Using the fact that $\partial_{\bar z} F_V(\varphi_t)=
\underset{k=1}{\overset{\infty}{\sum}} \frac{k}{\sqrt{k!}} \la (\varphi_t+\bar\varphi_t)^{\otimes k-1}|\otimes \11 \,V^{(k)}\in\Z$, we get
\bean
F_{V}(z+\varphi_t)&=& F_{R(t)}(z)+F_{V}(\varphi_t)+ \sum_{k=1}^\infty \frac{k}{\sqrt{k!}} \la(z+\bar z)\otimes (\varphi_t+\bar\varphi_t)^{\otimes k-1}, V^{(k)}\ra \\&&
+\sum_{k=2}^\infty \frac{C^2_k}{\sqrt{k!}} \la(z+\bar z)^{\otimes 2}\otimes
(\varphi_t+\bar\varphi_t)^{\otimes k-2}, V^{(k)}\ra\\
&=&F_{R(t)}(z)+F_{V}(\varphi_t)+2{\rm Re}\la z,\partial_{\bar z}F_V(\varphi_t)\ra+F_{V_2(t)}(z)\,,
\eean
where $F_{R(t)}$ is $\underset{n=3}{\overset{\infty}{\sum}} \la \frac{(z+\bar z)^{\otimes n}}{\sqrt{n!}}, \underset{k=n}{\overset{\infty}{\sum}}
\sqrt{\frac{n!}{k!}} C_k^n V^{(n)}_k\ra$.
This implies that the $t$-vector $\Theta'(t)$ is given by
$
\Theta'(t)= \frac{i}{\varepsilon}\, e^{i\frac{t}{\varepsilon}
H} e^{i\frac{\omega(t)}{\varepsilon}}
W(-i\frac{\sqrt{2}}{\varepsilon}\varphi_{t}) \;F_{R(t)}^{Wick} \;  U_{2}(t,0)\Psi,
$
if we choose $\omega$ such that 
$$
\partial_t \omega={\rm Re}\la\varphi_t, \partial_{\bar z}F_V(\varphi_t)\ra-F_V(\varphi_t)\,.
$$
This holds with
$$
\omega(t)=\ds\int_0^t \sum_{k=0}^\infty \frac{(k-2)}{2} \la \frac{(\varphi_s+\bar\varphi_s)^{\otimes k}}{\sqrt{k!}}, V^{(k)}\ra\, ds.
$$
Hence we conclude that
\bean
&& \hspace{-.4in} e^{i\frac{t}{\varepsilon}
H}e^{i\frac{\omega(t)}{\varepsilon}}
W(-i\frac{\sqrt{2}}{\varepsilon}\varphi_{t}) U_{2}(t,0)\Psi -W(-i\frac{\sqrt{2}}{\varepsilon}\varphi_{0})\Psi=\\
&&{}\hskip170pt\frac{i}{\varepsilon}\int_{0}^{t}
e^{i\frac{s}{\varepsilon}
H} e^{i\frac{\omega(s)}{\varepsilon}}
W(-i\frac{\sqrt{2}}{\varepsilon}\varphi_{s}) \;F_{R(s)}^{Wick} \;  U_{2}(s,0)\Psi \,~ds\,.
\eean
We observe that  $R(t)\in\underset{n\geq 3}{\oplus}\otimes_s^n\Z$ and proceed to estimate the right hand side.
So we have (for $t>0$)
$$
\left\|e^{-i \frac{t}{\varepsilon} H} W(-i\frac{\sqrt{2}}{\varepsilon}\varphi_0)\Psi-e^{i\frac{\omega(t)}{\varepsilon}}
W(-i\frac{\sqrt{2}}{\varepsilon}\varphi_{t})U_{2}(t,0)\Psi\right\|_{\F}\leq
\frac{1}{\varepsilon}\int_{0}^{t} \left\|\;F_{R(s)}^{Wick} \; U_{2}(s,0)\Psi\right\|_{\F} \,~ds\,,
$$
Now using the estimate (ii) of Lemma \ref{lem.domcomp}, with $0<\gamma<\alpha$,
we obtain
\bean
 \left\|\;F_{R(s)}^{Wick} \;  U_{2}(s,0)\Psi\right\|_{\F}\leq C(s)
 \varepsilon^{{\frac{3}{2}}} ||e^{\gamma\lambda^{\frac{N}{\varepsilon}}} R(s)||_{\Gamma_s(\Z)}\, ,
 \eean
such that   $C>0$ depending only on $(\alpha,\lambda)$ and
\bean
 C(s)= C \left[
||\sqrt{g_s(\frac{N}{\varepsilon})} \Psi||^2+ g'_s(0) \left|\int_0^s ||V_2(r)|| dr\right| ||\Psi||^2\right]^{\frac{1}{2}}\,,
\eean
where $g_t(r)=\underset{k=0}{\overset{\infty}{\sum}} e^{-\alpha_0\lambda^{k}} e^{ 2\sqrt{2} \lambda_0^k \int_0^t ||V_2(s)||ds}\, (r+1)^k$
 for $1<\lambda_0<\lambda$, $0<\alpha_0 \lambda^2<\alpha$ and $g'_t(r)=\frac{d}{dr}g_t(r)$.\\
 A similar estimate as in the proof of Lemma \ref{trans-sym} yields
\bean
||e^{\gamma\lambda^{\frac{N}{\varepsilon}}} R(s)||_{\Gamma_s(\Z)}&\leq& \sqrt{2} e^{4||\varphi_s||_\Z^2} \,||\sqrt{2}^{\frac{N}{\varepsilon}} \, e^{\gamma\lambda^{\frac{N}{\varepsilon}}}  V||_{\Gamma_s(\Z)}\\
&\leq& c \,e^{4||\varphi_s||_\Z^2} \,|| e^{\alpha\lambda^{\frac{N}{\varepsilon}}} V||_{\Gamma_s(\Z)}\,.
\eean
Hence there exist a $(\varepsilon,V,t)$-independent constant $c(t)>0$ such that
\bean
&&\left\|e^{-i \frac{t}{\varepsilon} H} W(-i\frac{\sqrt{2}}{\varepsilon}\varphi_0)\Psi-e^{i\frac{\omega(t)}{\varepsilon}}
W(-i\frac{\sqrt{2}}{\varepsilon}\varphi_{t})U_{2}(t,0)\Psi\right\|\leq
c(t)\sqrt{\varepsilon} ||e^{\alpha\lambda^{\frac{N}{\varepsilon}}} V|| \,,
\eean
with $C>0$ depending only on $(\alpha,\lambda)$ and
$$
c(t)= C \int_0^t e^{4||\varphi_s||_\Z^2} \left[
||\sqrt{g_s(\frac{N}{\varepsilon})} \Psi||^2+ g'_s(0) \left|\int_0^s ||V_2(r)|| dr\right| ||\Psi||^2\right]^{\frac{1}{2}} \; ds\,.
$$
Since $\Psi\in\D_f$ and $||\varphi_t||$ is bounded on compact intervals we see that the r.h.s is
finite.
\hfill\cqfd

\end{document}